\documentclass[letterpaper,journal]{IEEEtran}

\usepackage{amsmath,amsfonts}
\usepackage{algorithmic}
\usepackage{algorithm}
\usepackage{array}
\usepackage[caption=false,font=normalsize,labelfont=sf,textfont=sf]{subfig}
\usepackage{textcomp}
\usepackage{stfloats}
\usepackage{url}
\usepackage{verbatim}
\usepackage{graphicx}
\usepackage{cite}
\begin{document}

\title{STRIDe: Cross-Coupled \underline{ST}T-MRAM Enabling \underline{R}obust \underline{I}n-Memory-Computing for \underline{De}ep Neural Network Accelerators} 

\author{Imtiaz Ahmed and Sumeet Kumar Gupta,~\IEEEmembership{Senior Member,~IEEE}\vspace{-5mm}

\thanks{This work was supported in part by the Center for the Co-Design of Cognitive
Systems (CoCoSys), one of the seven centers sponsored by the Semiconductor Research Corporation (SRC) and DARPA under the Joint University Microelectronics Program 2.0 (JUMP 2.0), in part by Microelectronics Commons, in part by Raytheon, and in part by U.S. National Science Foundation (NSF).}
\thanks{The authors are with the Elmore Family School of Electrical and Computer
Engineering, Purdue University, West Lafayette, IN 47907, USA. (e-mail:
ahmed202@purdue.edu).}

}

\maketitle

\begin{abstract}
As deep neural network (DNN) models are growing exponentially in size, their deployment on resource-constrained edge platforms is becoming increasingly challenging. In-memory-computing (IMC) with non-volatile memories (NVMs) has emerged as a potential solution by virtue of its higher energy efficiency compared to standard DNN hardware platforms. Amongst various NVMs, STT-MRAM is highly promising owing to its high endurance and other benefits. However, their IMC implementation is challenging because of their inherently low distinguishability. This issue is exacerbated due to array non-idealities and process-variations, leading to poor IMC robustness and severe inference accuracy degradation. To address this problem, we propose STRIDe - STT-MRAM-based IMC leveraging cross-coupling action to boost the bitcell-level high-to-low current ratio to up to $\sim$$8000$. We propose two flavors of STRIDe designs, both offering robust IMC for inputs and weights  $\in$$\{-1,1\}$(XNOR-IMC) and $\{0,1\}$(AND-IMC). Our evaluations for STRIDe arrays show up to $3.86\times$ and $1.77\times$ sense margin (SM) improvement for XNOR-IMC and AND-IMC, respectively, and up to $27.6\%$ read disturb margin (RDM) improvement over standard MRAM-IMC designs. The enhanced robustness of STRIDe translates to near-software inference accuracies (considering crossbar non-idealities and process variations) for ResNet18 BNN and 4-bit DNN trained on CIFAR10 dataset. We observe  accuracy improvements of up to $\sim$$70\%$ (for BNN) and up to $\sim$$35\%$(for 4-bit DNN) over standard MRAM designs, albeit with some energy-area-latency penalty.
\end{abstract}

\begin{IEEEkeywords}
AND-IMC, Cross-coupled, Deep Neural Network (DNN), Distinguishability, In-Memory-Computing (IMC), Non-volatile memory(NVM), STT-MRAM, Tunneling Magnetoresistance (TMR), XNOR-IMC.\vspace{-1mm}
\end{IEEEkeywords}

\section{Introduction} \label{introduction}
\IEEEPARstart{D}{eep} Neural Networks (DNNs) have reshaped the landscape of artificial intelligence (AI), driving breakthroughs across tasks ranging from classification to generative modeling. However, the exponential increase in DNN model size has drastically elevated storage demands and the volume of energy-intensive memory–processor data transactions, resulting in the von Neumann bottleneck \cite{Strubell_Ganesh_McCallum_2020}. This leads to several challenges in their deployment on edge AI devices with stringent energy-area constraints. To address this, in-memory-computing (IMC) has emerged as a promising solution, which fuses storage and data-processing, enabling matrix-vector multiplications (MVM) directly within the crossbar memory macros\cite{9382915,8342235}. 

To further support the expanding scale of DNNs, quantization of DNN weights and inputs has been extensively studied for reducing storage cost and enhancing energy-efficiency. For instance, low-precision quantization of weights/inputs to 4-bits has shown high inference accuracies while reducing memory footprint \cite{choi2018pactparameterizedclippingactivation}. Even more aggressive quantization techniques have also been explored, such as in binary neural networks (BNN), in which weights/inputs are quantized to just 1-bit $\{-1,+1\}$. When used in conjunction with IMC, this ultra-low quantization drastically reduces storage, computation and data-communication demands while still retaining acceptable accuracy in certain inference tasks \cite{NIPS2016_d8330f85,8959407,8465935}.%whereas ternary neural network (TNN) uses the \{-1,0,+1\} quantization. Low precision representations like these 

Although traditional CMOS-based IMC implementation has been widely explored, the limited scalability of CMOS is struggling to keep up with the growing complexity of DNN models, resulting in an increasing gap between DNN demands and CMOS hardware performance\cite{thompson2022computationallimitsdeeplearning}. To bridge this gap, emerging non-volatile memories (NVMs) have emerged as promising solutions, offering high density and compute energy-efficiency through parallelized in-situ MVM operations\cite{8959407, 7047135, 8342235, Jung2022ACA,8768197,10268108}. The implementation of quantized-IMC with NVMs can yield synergistic performance improvements, pushing the boundaries of energy-area efficiency in edge devices. Various NVM technologies have been explored for IMC, such as phase-change memories (PCMs)\cite{7047135}, resistive RAMs (RRAMs)\cite{8565811}, and spintronic memories like spin-transfer-torque magnetic RAMs (STT-MRAM)\cite{10.1145/2463585.2463589}- all with their own pros and cons. Among these, STT-MRAM is a particularly exciting memory candidate due to its high endurance, long retention time, and CMOS-compatible process and programming voltages\cite{8675492}. However, MRAM-based IMC implementation is challenging because MRAMs suffer from low tunneling magnetoresistance (TMR), resulting in poor distinguishability between logic states '0' and '1'\cite{8976130}. As a result, large '0' currents are generated, potentially summing up to a false '1' current and introducing compute errors. This concern is further exacerbated by the circuit non-idealities like parasitic resistances in the crossbar arrays and device-to-device variations, as they cause deviation from ideal current and increase the probability of IMC error. Despite a maximum room temperature TMR of 631\% being reported \cite{10.1063/5.0145873}, typical MRAM designs are restricted to smaller TMR \cite{8976130}. 

Moreover, IMC-robustness also depends on the bit-cell topologies. Existing MRAM-based IMC designs include AND-based 1T-1MTJ designs with \textit{unsigned} weights/inputs ($\in$$\{0,1\}$) \cite{8241447,8675209} and XNOR-based 2T-2MTJ designs with \textit{signed} weights/inputs ($\in$$\{-1,1\}$)\cite{9762321}. Though the latter trades off crossbar array area for higher robustness compared to the former (details later), both designs are, in general, susceptible to the low distinguishability problem, circuit non-idealities and process-variations. Therefore, to enable robust MRAM-IMC implementation while also reaping the benefits of their other memory features, it is imperative to explore distinguishability-enhancement strategies for STT-MRAMs. Now, implementing robust MRAM-IMC and exploring non-ideality mitigation techniques have been active research pursuits, involving circuit techniques \cite{7551379,8714817,9678974,Jung2022ACA}, device-algorithm co-design for IMC optimization\cite{9502492}, and others. Despite their own sets of strengths and limitations, most of these techniques do not address the fundamental concern of the low distinguishability of MRAM. Hence, new MRAM-IMC designs need to be explored to enhance robustness at the bitcell level. 

To that end, earlier we have proposed CREST-CiM, a robust differential MRAM-based XNOR-IMC design for BNNs that leverages cross-coupling action between two STT-MRAM branches to \textit{locally} boost the ratio of currents corresponding to logic '1' and '0' (I\textsubscript{H}/I\textsubscript{L}) \cite{11133247}. In this work, we take this cross-coupling principle further to push the limits of robust MRAM-IMC by proposing STRIDe (Cross-Coupled STT-MRAM Enabling Robust IMC for DNN Accelerators). We propose two flavors of cross-coupled designs, STRIDe-I and STRIDe-II, show how they enhance distinguishability at the bitcell level, implement robust XNOR-IMC and AND-IMC, and achieve high inference accuracy for BNNs and higher precision (such as 4-bit) DNNs. The contributions of this work are as follows:  

\begin{itemize}
\item We propose STRIDe-I and II, two STT-MRAM based IMC designs where each bitcell stores binary weights using a pair of magnetic tunnel junctions (MTJ), and achieves significant bitcell-level distinguishability-enhancement through cross-coupling action between the two MTJ branches. The maximum achieved I\textsubscript{H}/I\textsubscript{L} ratios are approximately 8156 for STRIDe-I bitcell and 3932 for STRIDe-II bitcell (a significant boost from the I\textsubscript{H}/I\textsubscript{L}$\sim$$5$ for standard STT-MRAM used in this work).  

\item We design 64$\times$64 STRIDe IMC-crossbar arrays with circuit non-idealities, and achieve significant improvements in sense margin (SM) and read disturb margin (RDM) over standard MRAM-IMC designs for both XNOR and AND-based IMC. The SM enhancement also makes STRIDe designs process-variation tolerant.

\item We deploy ResNet18 BNN and 4-bit DNN inference workloads (trained on CIFAR10) on STRIDe and demonstrate near-software inference accuracy owing to their IMC-robustness and process-variation tolerance. This represents an accuracy increase by $\sim$$35\%$ to $\sim$$70\%$ for BNN and $\sim$$5\%$ to $\sim$$37\%$ for 4-bit DNN over previous standard MRAM-IMC designs, at the cost of some energy-latency-area overhead.
\end{itemize}

\section{Preliminaries and Related Works} \label{relatedworks}

\subsection{Magnetic Tunnel Junction (MTJ) Based STT-MRAM}
\begin{figure}[t!]
\centering
\includegraphics[width=3.4in]{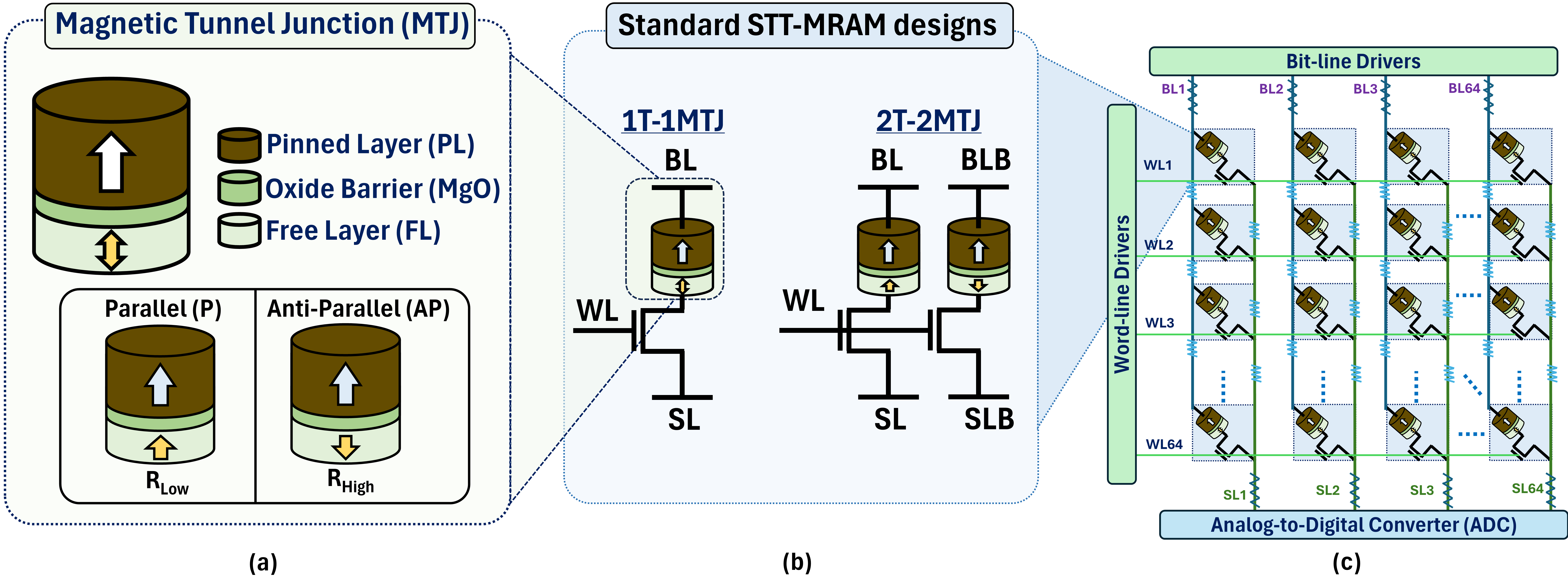}\vspace{-2mm}
\caption{Magnetic Tunnel Junction (MTJ) based STT-MRAM. (a)Layers of MTJ. PL and FL are insulated with oxide (MgO) layer. (b)1T-1MTJ and 2T-2MTJ based STT-MRAM bitcells.(c)STT-MRAM based IMC Crossbar arrays.\vspace{-4mm}}
\label{Figure: MTJ}
\end{figure}

STT-MRAMs utilize non-volatile memory elements based on MTJ which contains two ferromagnet layers (pinned layer, PL and free layer, FL) separated by a thin insulating tunneling oxide layer, usually MgO(Fig. \ref{Figure: MTJ}a). PL has a fixed magnetization orientation. In contrast, the magnetization orientation of FL can be switched with spin-current induced torque. Based on the relative magnetization orientation of PL and FL, MTJs can retain two stable memory states- namely, the parallel state (P) with PL and FL oriented in the same direction, and the anti-parallel state (AP) with PL and FL oriented in the opposite direction. The former exhibits low resistance ($R_{P}$=$R_{LOW}$) while the latter has a high resistance ($R_{AP}$=$R_{HIGH}$). A key figure of merit for the distinguishability between these two states is the tunneling magnetoresistance, $\text{TMR}$=$\frac{R_{AP}-R_{P}}{R_{P}}$$\times$$100\% $. Higher TMR means more distinguishable memory states. However, MTJs inherently have low TMR, limiting their distinguishability and making it challenging to implement robust MRAM-IMC. \vspace{-2mm} 

\subsection{Standard MRAM-based IMC Designs}

Several works have explored MRAM-IMC,  with the standard 1T-1MTJ topology (Fig. \ref{Figure: MTJ}b) being amongst the most dense designs. Here, both weights and inputs are denoted as binary \{$0,1$\} using only one MTJ per bit-cell, leading to a high density design. If 4-bit precision IMC needs to be implemented with this, the 4-bit weights are stored in 4 crossbar arrays, each storing a bit-slice of 1. Also, the 4-bit inputs are streamed in 4 cycles as binary wordline(WL) voltages. The scalar product of the inputs and weights (and hence, the MVM output) is obtained via their bit-wise AND. 
  
On the other hand, binary neural network (BNN) has both weights and inputs denoted as \{$-1,+1$\}. Hence, bitwise XNOR-operation between input and weight corresponds to their scalar product. There are two standard strategies to achieve this. One way is to utilize the 1T-1MTJ based NAND-Net design, with inputs and weights converted to \{$0,1$\} domain and output post-processed to get the XNOR-IMC output \cite{8675209}. Another way is to design custom bit-cells with two MTJs per bit-cell storing complementary weights to encode $-1$ or $+1$ (e.g. a differential 2T-2MTJ structure, see Fig. \ref{Figure: MTJ}b)\cite{9762321}. This approach can perform in-situ XNOR operation, and provides better robustness than 1T-1MTJ, albeit with higher array area. An important point to note is that in the IMC macros, crossbar-array area takes up a small fraction of the entire memory macro with all peripherals (especially analog-to-digital converters or ADCs) considered \cite{10642976}. This makes it a common approach in literature to trade-off array area for achieving more robust MRAM-IMC \cite{9762321,Jung2022ACA}. Moreover, although the 1T-1MTJ design requires lower array area, it incurs additional (but mild) hardware cost due to the additional peripheral circuits required for transforming inputs from \{$-1,+1$\} to \{$0,1$\} and the IMC array outputs from \{$0,1$\} to \{$-1,+1$\} (for instance, adder trees for dynamic calculation of the sum of inputs \cite{8675209}).

\subsection{Challenges of MRAM-IMC: Low TMR and Circuit Non-Idealities}

Standard MRAM-IMC designs suffer primarily due to low TMR, or poor high-to-low current ratio (I\textsubscript{H}/I\textsubscript{L}) of MTJs. This is further exacerbated by the driver/sink/parasitic resistances in the crossbar and other non-idealities such as process variations. Due to low TMR, logic '0' state has a large I\textsubscript{L} current, causing poor distinguishability between logic states '0' and '1'. This becomes particularly problematic in IMC, which requires the assertion of multiple wordlines, and just a few '0' state currents can add up to produce a false '1' current. In addition, circuit non-idealities further worsen the robustness issues already introduced by low TMR, leading to computational errors. As currents from multiple bit-cells accumulate, the bit-lines carry much larger currents compared to standard memory-read operation. This causes large IR drops in the non-ideal resistances and results in deviations in the output currents from the ideal/expected values. As a combined effect of low TMR and non-ideal crossbar behavior, sense margin (SM) gets severely degraded, which drastically increases the probability of overlaps between neighboring output states. These effects become even worse under process-induced variations, significantly impairing the computational robustness.\vspace{-3mm}

\subsection{Existing Strategies towards Robust MRAM-IMC}
 
Given the challenges of MRAM-IMC, enhancing its robustness has been an active research endeavor, approached from multiple fronts such as technology innovations, robust circuit techniques, device-circuit co-design and others. There have been multiple device-level efforts to come up with MTJs with high TMR\cite{10.1063/1.2976435,10.1063/5.0145873}. While promising, these innovations are not mature yet and require systematic investigation before their deployment. Additionally, several novel bit-cell designs have been proposed to circumvent the low distinguishability issue, with a common approach being trading-off array-area for improved robustness, as noted earlier. The 2T-1MTJ bitcells in \cite{PATEL2014133} provides enhanced I\textsubscript{H}/I\textsubscript{L} ratio for standard memory read, but they may be prone to elevated read-disturb probability and are yet to be explored for IMC. Bitcell design with decoupled read-write path in \cite{9502492} allows for IMC-specific optimization. However, this still suffers from the low TMR problem. Differential 2T-2MTJ bit-cells have shown promise with enhanced robustness due to the cancellation of the common-mode noise \cite{10642976}. Further, The resistance-sum approach utilizing 2T-2MTJ bitcells is an innovative strategy with high energy efficiency, although it is limited to time-based sensing (which is not as fast as current sensing)\cite{Jung2022ACA}.

Furthermore, multiple circuit techniques have also been explored to enhance IMC robustness, which can be used in combination with one another. As noted earlier, large I\textsubscript{L} produced by standard MRAM-bitcells is the primary culprit for poor robustness. To mitigate this, a dummy column with all '0' weights stored has been used in \cite{7551379,10933557}, which shares its input with the regular crossbar array. The dummy column current is subtracted from the real column output currents to mitigate the effect of false '1's. But, this correction is not perfect due to crossbar non-idealities impacting real columns and dummy column differently. Partial wordline activation (PWA) is another effective method, where only a subset of the wordlines are asserted to reduce IR drops, albeit with higher latency \cite{8714817,10268108}. An additional benefit of this approach is the reduction of precision requirement for ADC, lowering the ADC energy/area costs. Dynamic latching of 1T-1MTJ array has achieved significant TMR-magnification; however, the magnification is heavily dependent on peripheral circuitry and suffers from reduced parallelism \cite{9678974}. 

Despite having their own strengths and limitations, these approaches offer workarounds to the fundamental low-distinguishability problem rather than solving this problem itself at the bitcell level. To address this limitation, our proposed STRIDe designs directly target the issue of low I\textsubscript{H}/I\textsubscript{L} ratio  and enhances distinguishability at the very bitcell level. This maximizes column-parallelism while achieving significant enhancement in IMC robustness. Besides, STRIDe can be used in conjunction with other non-ideality mitigation techniques for further performance improvements (details later).

\begin{figure*}[t]
\centering
\includegraphics[width=0.9\textwidth]{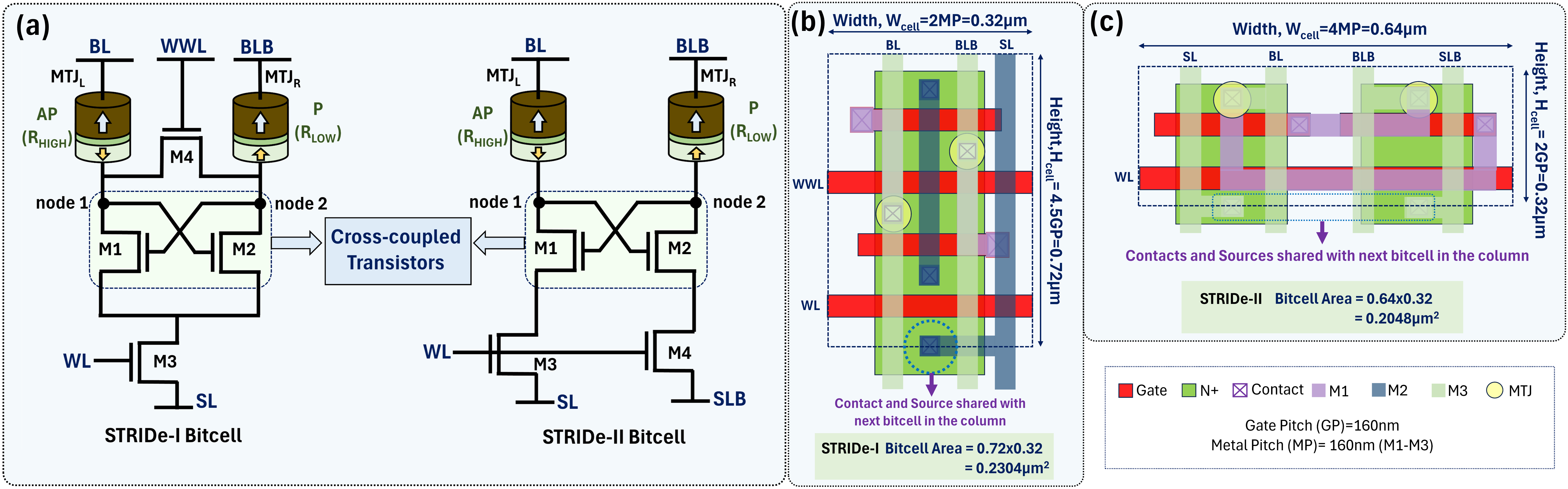}\vspace{-1mm}
\caption{(a) Circuit diagram for STRIDe-I and II bitcells, (b) STRIDe-I bitcell layout (45nm node), modified from \cite{11133247}, (c) STRIDe-II bitcell layout.\vspace{-7mm}}
\label{Figure: bitcells}
\end{figure*}

\section{STRIDe: Cross-Coupled STT-MRAM with Enhanced Distinguishability} \label{stride}
In this section, we introduce the working principle of STRIDe-I and II bitcells, their write/read operations, and the analysis of cross-coupling-enhanced I\textsubscript{H}/I\textsubscript{L} ratio. Let us begin with the description of our simulation framework.\vspace{-3mm}

\subsection{Simulation Framework}

For our bitcell design and simulations, we use perpendicular magnetic anisotropy (PMA) MTJ models from \cite{nanoHUB.org16}. The parameters listed in Table \ref{table1}, calibrated against experimental data, have been taken from \cite{9210520,worledge2011spin}. The Landau-Lifshitz-Gilbert-Slonczewski model is used to characterize the FL magnetization switching behavior, thereby capturing the dynamics of the MTJ \cite{6035047}. Also, the non-equilibrium Green’s function (NEGF) model is utilized to model the MTJ resistance \cite{6035047}. The MTJ-TMR is \(\sim \)400\% to 450\% following the works in \cite{9502492,10.1063/5.0145873} to enhance the IMC robustness of the baseline MRAM-designs for presenting a fair comparison in the subsequent sections. For the transistors, we utilize the predictive technology models corresponding to the 45nm technology node \cite{4418914,4231502}.

\begin{table}[!b]
\scriptsize
\caption{Device and Circuit Parameters for HSPICE Simulations}
\begin{center}
\begin{tabular}{p{140pt}p{80pt}}
\hline
Parameters & Value \\ \hline
Free Layer Dimension, $W\times L$ ($nm^{2}$) & $60\times 60$\\ 
FL Thickness, $T_{M}$ $(nm)$  &  $1$\\ 
MgO Thickness, $t_{ox}$ $(nm)$ & $1.3$\\ 
Saturation Magnetization, $M_{s}$ $(emu/cm^{3})$
& $865$ \cite{9210520}\\ 
Uniaxial Anisotropy Density, $K_{u}$ $(erg/cm^{3})$ & $9.66\times 10^{5}$ \cite{worledge2011spin}\\
Energy Barrier, $E_{B}(k_{B}T)$  & $64$\\ 
Damping Coefficient, $\alpha $ & $0.008$\\ 
Gyromagnetic Ratio, $\gamma$ $(MH_z/Oe)$ & $17.6$\\ 
Wire Resistance, $R_w$$(\Omega /\mu m)$ & $3.3$ \cite{moon2008process} \\ 
Driver Resistance, $R_D$ $(\Omega)$ & $250$\\ \hline
\end{tabular}
\vspace{-20pt}
\label{table1}
\end{center}
 \end{table}

\subsection{STRIDe Bitcells} \label{bitcells}
Fig. \ref{Figure: bitcells}a shows the schematics of the STRIDe bitcells. Both bitcells contain two MTJs (MTJ\textsubscript{L} and MTJ\textsubscript{R}) storing complementary weights (similar to previous 2T-2MTJ designs \cite{9762321}). For XNOR-IMC, weight = $+1$ is encoded as MTJ\textsubscript{L} in the parallel (P) state and MTJ\textsubscript{R} in the anti-parallel (AP) state. Weight = $-1$ is encoded as AP stored in MTJ\textsubscript{L} and P stored in MTJ\textsubscript{R} (details later). For AND-IMC, weight=$1$ corresponds to  MTJ\textsubscript{L} and MTJ\textsubscript{R} in the parallel (P) and anti-parallel (AP) states, respectively, while weight=0 is the other way around. In both bitcells, the MTJs are cross-coupled via transistors M1 and M2. In STRIDe-I bitcell, a common access transistor M3 is connected to the sources of M1 and M2, which is controlled by the wordline (WL), as shown in Fig. \ref{Figure: bitcells}a. Also, a write access transistor M4 is connected to the drains of M1 and M2, controlled by write wordline (WWL). On the contrary, STRIDe-II bitcell has two access transistors M3 and M4 (connected to the sources of M1 and M2, respectively), both controlled by the same wordline (WL). Unlike the STRIDe-I bitcell, STRIDe-II has no separate write transistor. For both bitcells, PL of MTJ\textsubscript{L} and MTJ\textsubscript{R} are connected to the bit-lines, BL and BLB, respectively.

Fig. \ref{Figure: bitcells}b and \ref{Figure: bitcells}c show the bitcell layouts of STRIDe-I and II bitcells, respectively, following the design rules and predictive models for 45nm technology node \cite{4418914,4231502}. The transistors have been optimized with contact sharing for making the layouts compact. For STRIDe-I, WL and WWL are routed horizontally, while BL, BLB, and SL are routed vertically on M2 and M3 metal layers (Fig. \ref{Figure: bitcells}b). As for STRIDe-II, WL is routed horizontally while BL, BLB, SL, and SLB are routed vertically on M3 metal layer (Fig. \ref{Figure: bitcells}c). Note that, the bitcell area of STRIDe-II is \(\sim \)$11\%$ smaller than that of STRIDe-I.

Let us discuss the write and read operations of these bitcells.

\subsubsection{Write Operation}

\begin{figure}[t!]
\centering
\includegraphics[width=1.5in]{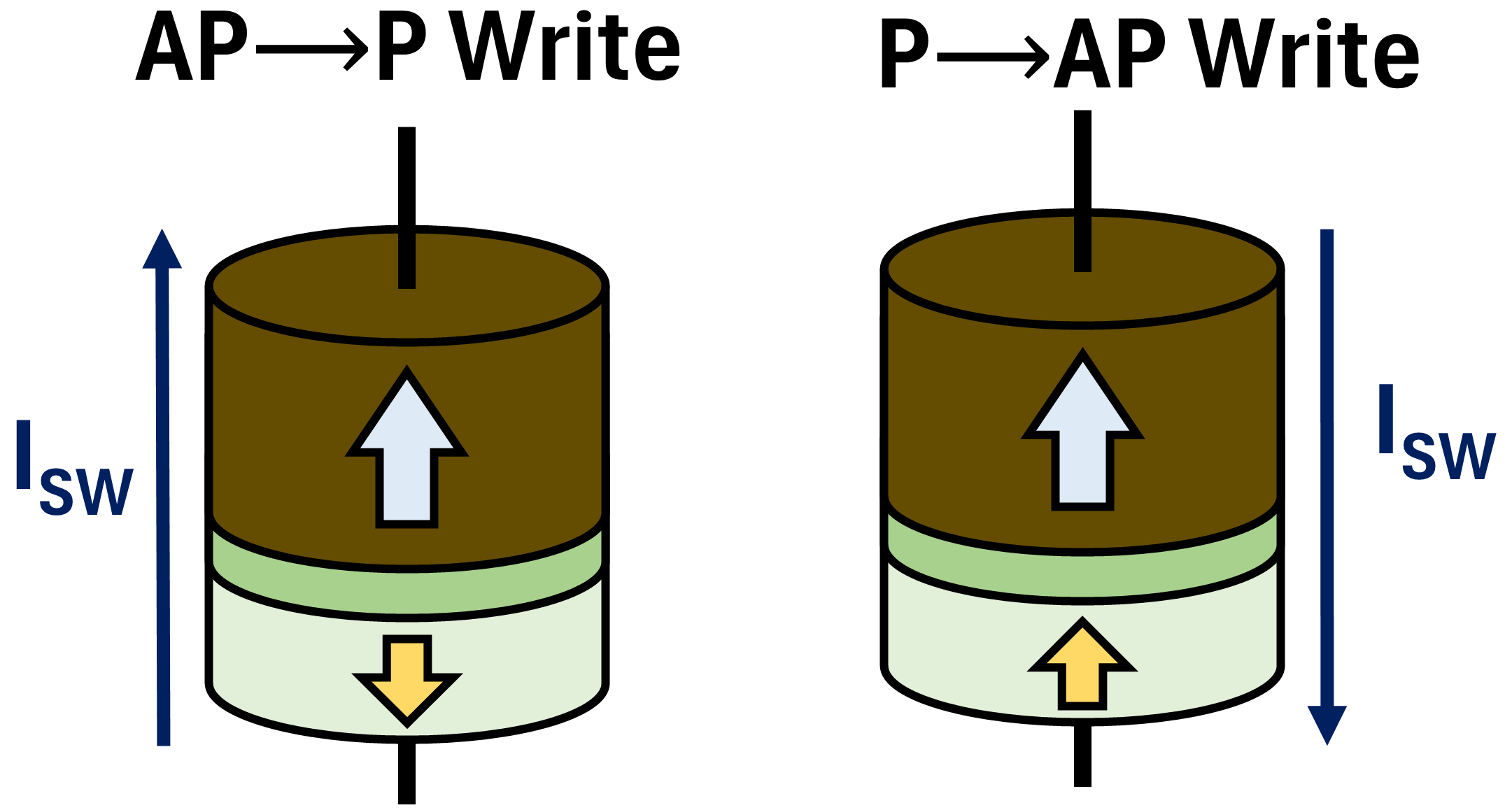}\vspace{-2mm}
\caption{Direction of switching current required to write into MTJ\cite{11133247}. \vspace{-4mm}}
\label{Figure: switch current direction}
\end{figure}

\begin{figure}[t!]
\centering
\includegraphics[width=0.49\textwidth]{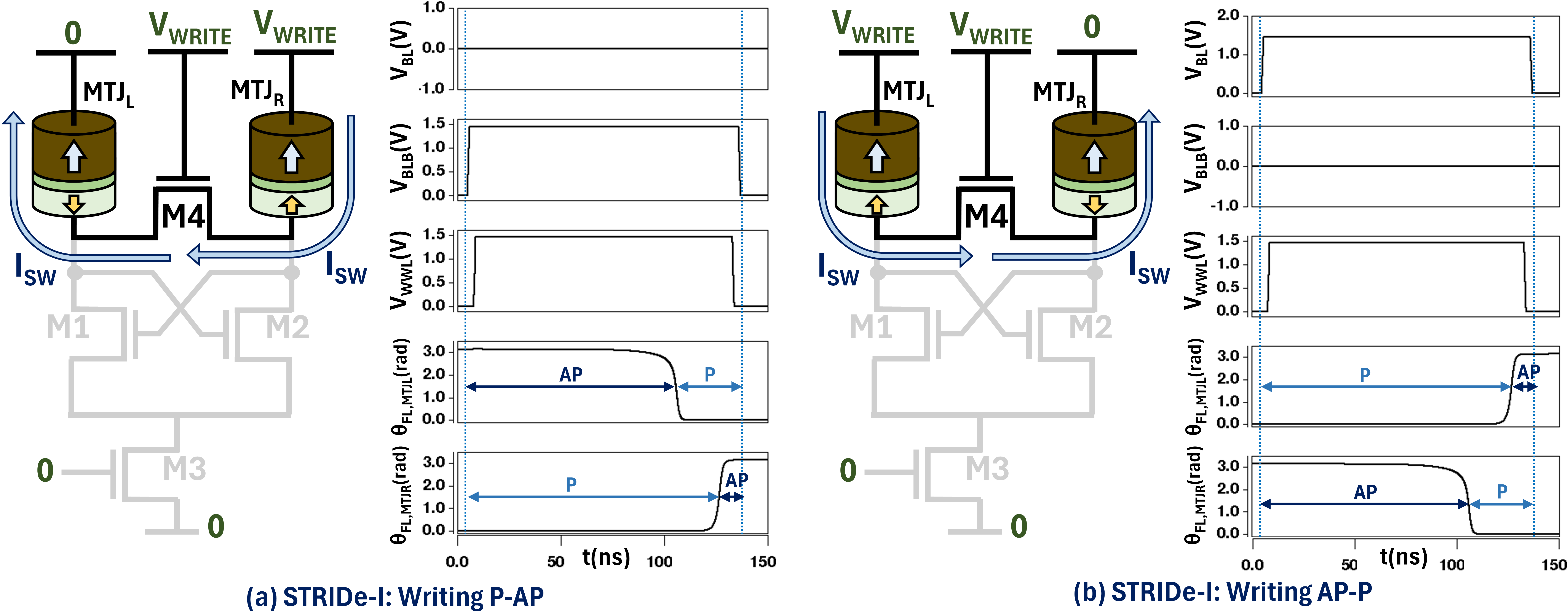}\vspace{-2mm}
\caption{Write Operation in STRIDe-I for (a) P-AP programming and (b) AP-P programming. WWL is asserted, turning M4 on and allowing controlled flow of switching current depending on BL/BLB voltages.\vspace{-4mm}}
\label{Figure: stride1write}
\end{figure}

The current direction through MTJ required for AP\textrightarrow P and P\textrightarrow AP switching is shown in Fig. \ref{Figure: switch current direction}. For the STRIDe-I bitcell, this switching current can be controlled with WWL. To program MTJ\textsubscript{L} to P and MTJ\textsubscript{R} to AP, we apply V\textsubscript{WRITE}/0 to BLB/BL while asserting WWL and keeping WL at 0 (Fig. \ref{Figure: stride1write}a). As current flows from BLB to BL, MTJ\textsubscript{R} switches to AP (due to current from PL to FL, Fig. \ref{Figure: switch current direction}) and MTJ\textsubscript{L} switches to P (due to current from FL to PL). As both the MTJs gets programmed simultaneously, STRIDe\textemdash I requires only one write-cycle. Similarly, to program MTJ\textsubscript{L} to AP and MTJ\textsubscript{R} to P, the current direction is reversed by applying V\textsubscript{WRITE}/0 on BL/BLB (Fig. \ref{Figure: stride1write}b). As BL and BLB are routed along the column, and WWL along the row, in the memory array, this enables simultaneous write in multiple cells of a row over a single cycle. Unlike the standard STT-MRAM with 1-MTJ and 1-transistor in the write path, STRIDe-I has 2-MTJs and 1-transistor in the write path. resulting in $1.44$x higher write latency and $1.39$x higher write energy compared to standard 1T-1MTJ for V\textsubscript{WRITE}=$1.45V$. However, since the target application for STRIDe is DNN accelerators using weight stationary architectures, the writes are quite infrequent, while MVM computation is the most dominant operation. Hence, to enhance the robustness and parallelism of MVM-IMC, trading-off write efficiency is reasonable (similar to some previous designs \cite{10.1063/9.0000378,thakuria2024sitecimsignedternary}).

\begin{figure}[t!]
\centering
\includegraphics[width=0.49\textwidth]{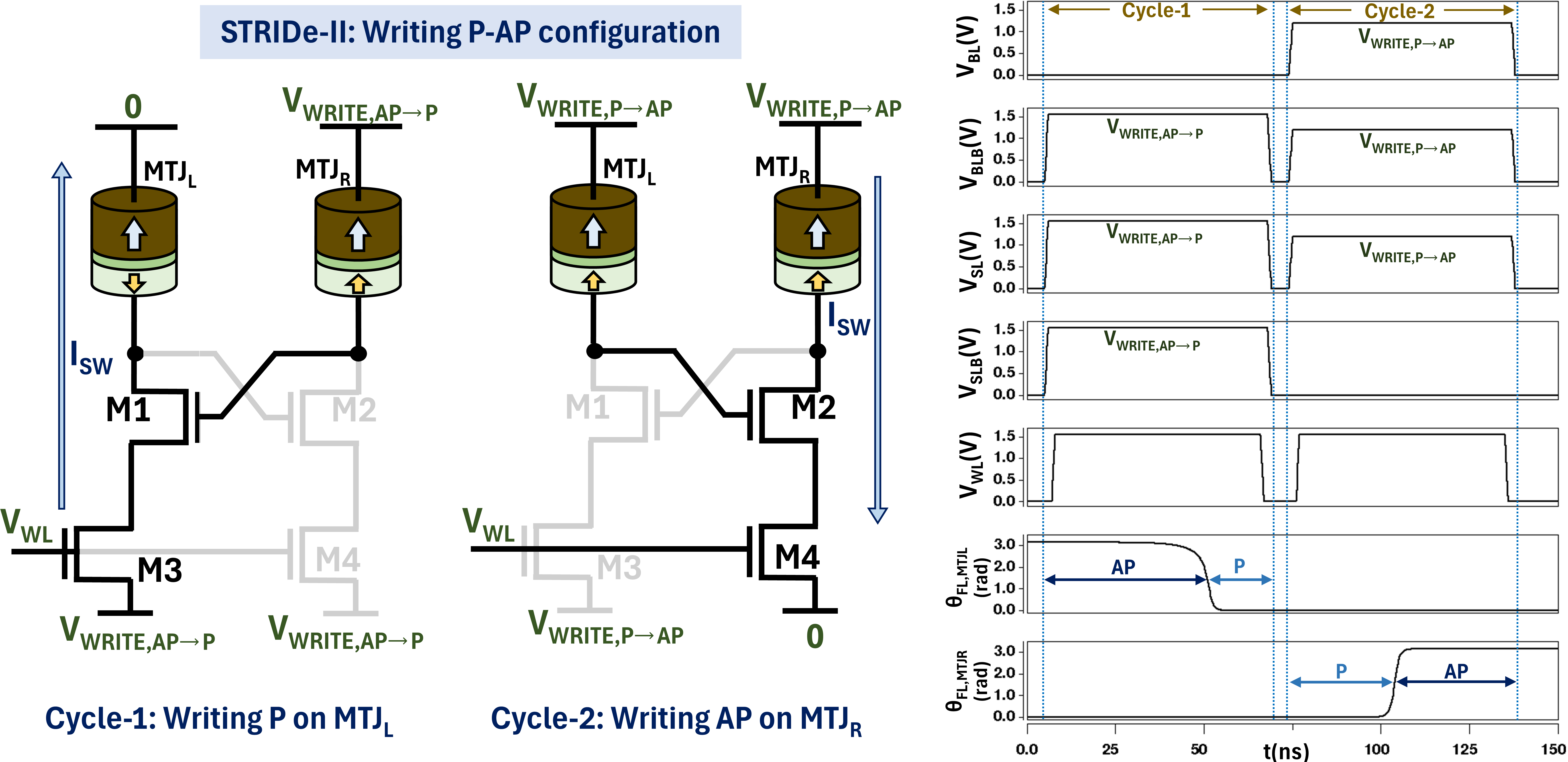}\vspace{-2mm}
\caption{Two-cycle write of P-AP in STRIDe-II. MTJ\textsubscript{L} and MTJ\textsubscript{R} are written during first and second cycle, respectively. Direction of current flow is controlled with appropriate terminal bias voltages while keeping WL asserted.\vspace{-5mm}}
\label{Figure: stride2write}
\end{figure}

\begin{figure}[b!]
\centering
\includegraphics[width=0.47\textwidth]{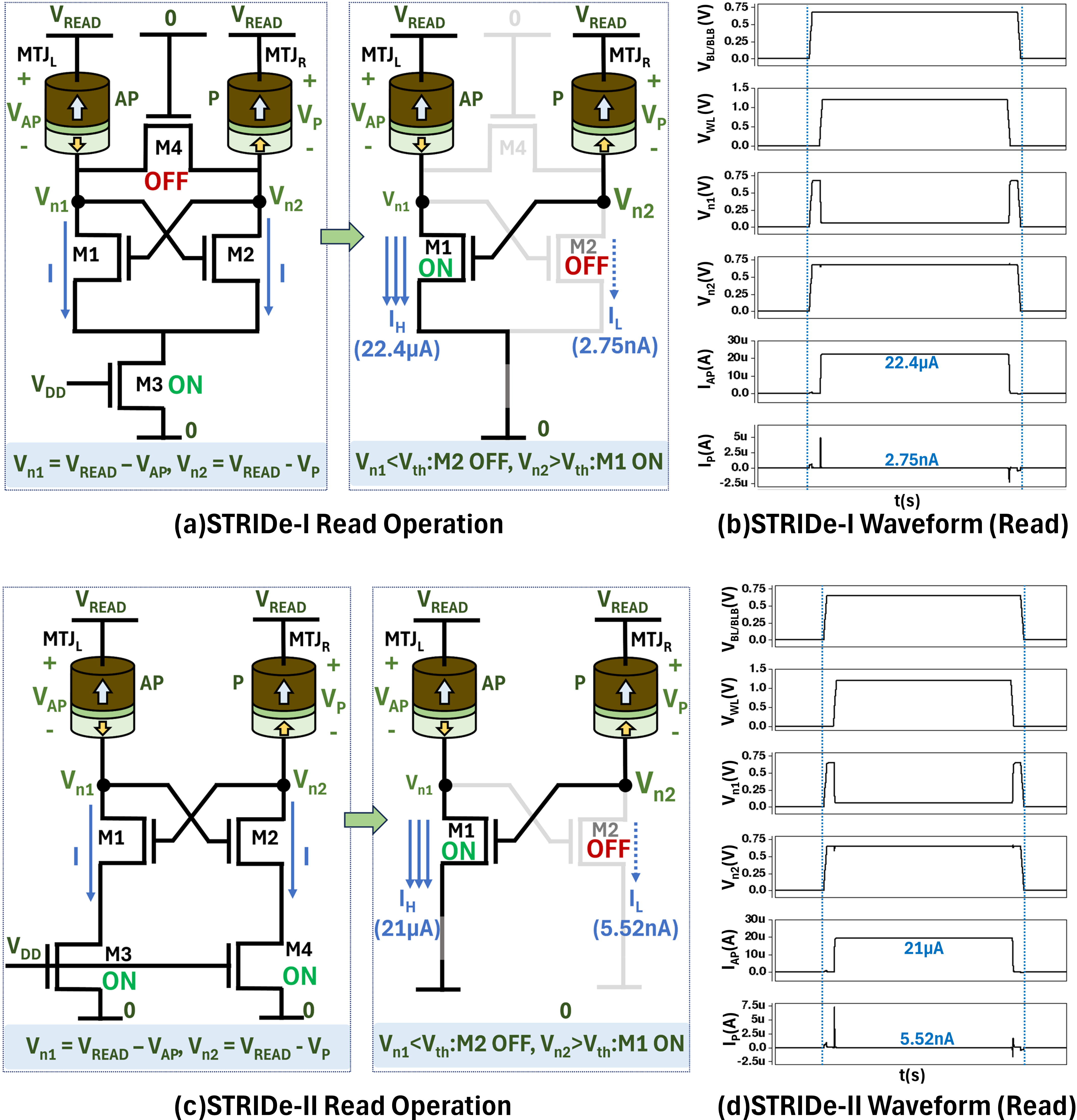}\vspace{-2mm}
\caption{Read operation of STRIDe-bitcells for AP-P weight. (a) For STRIDe-I, V\textsubscript{READ} is applied to BL/BLB with WL asserted. Internal node voltage at AP branch(V\textsubscript{n1}) is smaller than at P branch (V\textsubscript{n2}), turning M2 OFF and M1 ON. This cross-coupling action reduces I\textsubscript{P} and boosts I\textsubscript{AP}/I\textsubscript{P} ratio (e.g. I\textsubscript{H}/I\textsubscript{L} ratio). (b) Read operation waveform for STRIDe-I bitcell. (c) Similarly for STRIDe-II, V\textsubscript{READ} is applied to BL/BLB, and cross-coupling action takes place. (d) Read operation waveform for STRIDe-II bitcell.\vspace{-4mm}}
%Current vs V\textsubscript{READ} (left) and I\textsubscript{H}/I\textsubscript{L} vs V\textsubscript{READ}(right) for STRIDe-II bitcell. \vspace{-4mm}}
\label{Figure: read}
\end{figure}

Unlike STRIDe-I, STRIDe-II bitcell does not have a separate write-control transistor and requires a two-cycle write. Let us take an example of programming MTJ\textsubscript{L} to P and MTJ\textsubscript{R} to AP. During the first cycle, we apply V\textsubscript{WRITE}=$1.55V$ to WL, BLB, SL, and SLB, while keeping BL at 0 to write P on MTJ\textsubscript{L}(Fig. \ref{Figure: stride2write}). In the second cycle, V\textsubscript{WRITE}=$1.2V$ is applied to WL, BL, BLB, and SL, while keeping SLB at 0, for writing AP on MTJ\textsubscript{R}. Similarly, for programming MTJ\textsubscript{L} to AP and MTJ\textsubscript{R} to P, we apply V\textsubscript{WRITE}=$1.2V$ to WL, BL, BLB, and SLB, while keeping SL at 0 during the first cycle (writing AP into MTJ\textsubscript{L}), whereas in the second cycle, V\textsubscript{WRITE}=$1.55V$ is applied to WL, BL, SL, and SLB, while keeping BLB at 0 (writing P into MTJ\textsubscript{R}). Such a two-cycle write has a $1.64$x higher write latency and $1.54$x higher write energy compared to the 1T-1MTJ bitcell. However, as mentioned before, it is worthwhile to trade-off write efficiency for enhanced IMC robustness for weight stationary DNN inference.

\subsubsection{Read Operation}
Fig. \ref{Figure: read} demonstrates the read operation of STRIDe-I and II bitcells. During read (or IMC) operation of both bitcells, BL and BLB are driven to V\textsubscript{READ} and WL is asserted with V\textsubscript{DD}=$1.2V$ to turn on the access transistors (M3 for STRIDe-I, M3 and M4 for STRIDe-II). Also, for STRIDe-I, SL and WWL are kept at 0, and for STRIDe-II, both SL and SLB are kept at 0. This results in the scenarios as shown in Fig. \ref{Figure: read}a and Fig. \ref{Figure: read}c. 

To understand the I\textsubscript{H}/I\textsubscript{L} boost during read operation, let us consider MTJ\textsubscript{L} to be in the AP state (R\textsubscript{HIGH}) and MTJ\textsubscript{R}\ in the P state (R\textsubscript{LOW}). As BL/BLB are initially at 0, node-1 and node-2 are discharged in the beginning, meaning initial V\textsubscript{n1} = V\textsubscript{n2} = 0. As V\textsubscript{READ} is applied to both BL-BLB, node-1 and node-2 start getting charged through their respective MTJs. However, due to the resistance difference of the MTJs, node-2 charges faster than node-1. As the MTJ branches are cross-coupled and node-2 drives the gate of M1, M1 turns ON first with the application of proper V\textsubscript{READ}, leading to a high current (a few tens of \textmu A) in the left (AP) branch. Also, with M1 ON, V\textsubscript{n1} is pulled down below the threshold voltage of M2, driving M2 into OFF state. With M2 OFF, a very low current in the range of a few nA flows through the right (P) branch. Moreover, V\textsubscript{n2} reaches almost V\textsubscript{READ} in the steady state. As a result, the ON state of M1 (and the resulting high current in the AP branch) is reinforced  (Fig. \ref{Figure: read}a,c). The waveforms associated with read-operation are demonstrated in Fig. \ref{Figure: read}b,d. Similarly, if MTJ\textsubscript{L} and MTJ\textsubscript{R}\ store P and AP, respectively, this results in a very low current on the left (P) branch and high current on the right (AP) branch. Thus, the cross-coupling significantly enhances the ratio of high and low currents (I\textsubscript{H}/I\textsubscript{L}) due to ON current (I\textsubscript{H}) flowing in the AP branch, and OFF current (I\textsubscript{L}) flowing in the P branch.

\begin{figure}[b!]
\centering
\includegraphics[width=0.49\textwidth]{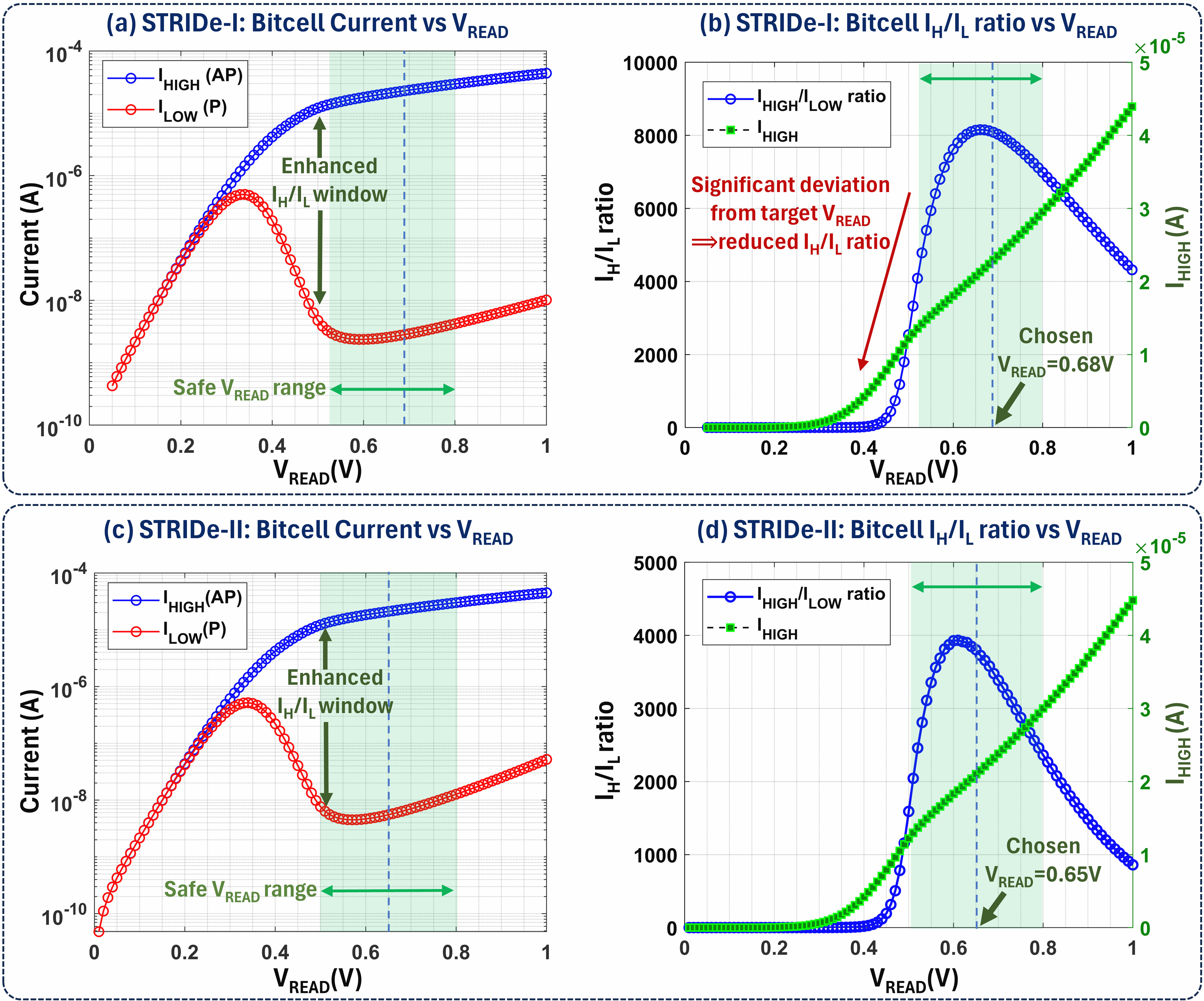}\vspace{-2mm}
\caption{Windows of enhanced distinguishability with STRIDe. (a)Current vs V\textsubscript{READ} and (b)I\textsubscript{H}/I\textsubscript{L} vs V\textsubscript{READ} for STRIDe-I. (c)Current vs V\textsubscript{READ} and (d)I\textsubscript{H}/I\textsubscript{L} vs V\textsubscript{READ} for STRIDe-II.\vspace{-4mm}}

\label{Figure: distinguishability_enhancement}
\end{figure}

Currents from the two branches can be sensed at BL and/or BLB for STRIDe-I and at SL and/or SLB for STRIDe-II. For XNOR-IMC, currents are sensed from both the branches, and subtracted using analog current-subtractor to obtain the output current, given by $I_{OUT}=I_{BLB}-I_{BL}$ for STRIDe-I and $I_{OUT}=I_{SLB}-I_{SL}$ for STRIDe-II. Essentially, $I_{OUT}$ is positive for P-AP and negative for AP-P state (details in section \ref{crossbar_design}). For AND-IMC, currents are sensed only at BLB (for STRIDe-I) or SLB (for STRIDe-II), and the output current is given by $I_{OUT}=I_{BLB}$ for STRIDe-I and $I_{OUT}=I_{SLB}$ for STRIDe-II. These output currents are then passed through ADCs to get the digital output. (It is worth mentioning that for XNOR-IMC, an alternate method would be to digitize the BL/SL and BLB/SLB currents first using two ADCs and then subtract using a digital subtractor \cite{11202915}. The choice would depend on the relative costs of the ADC and the subtractor. Here, we focus only on the former approach.)

\begin{figure}[t!]
\centering
\includegraphics[width=0.45\textwidth]{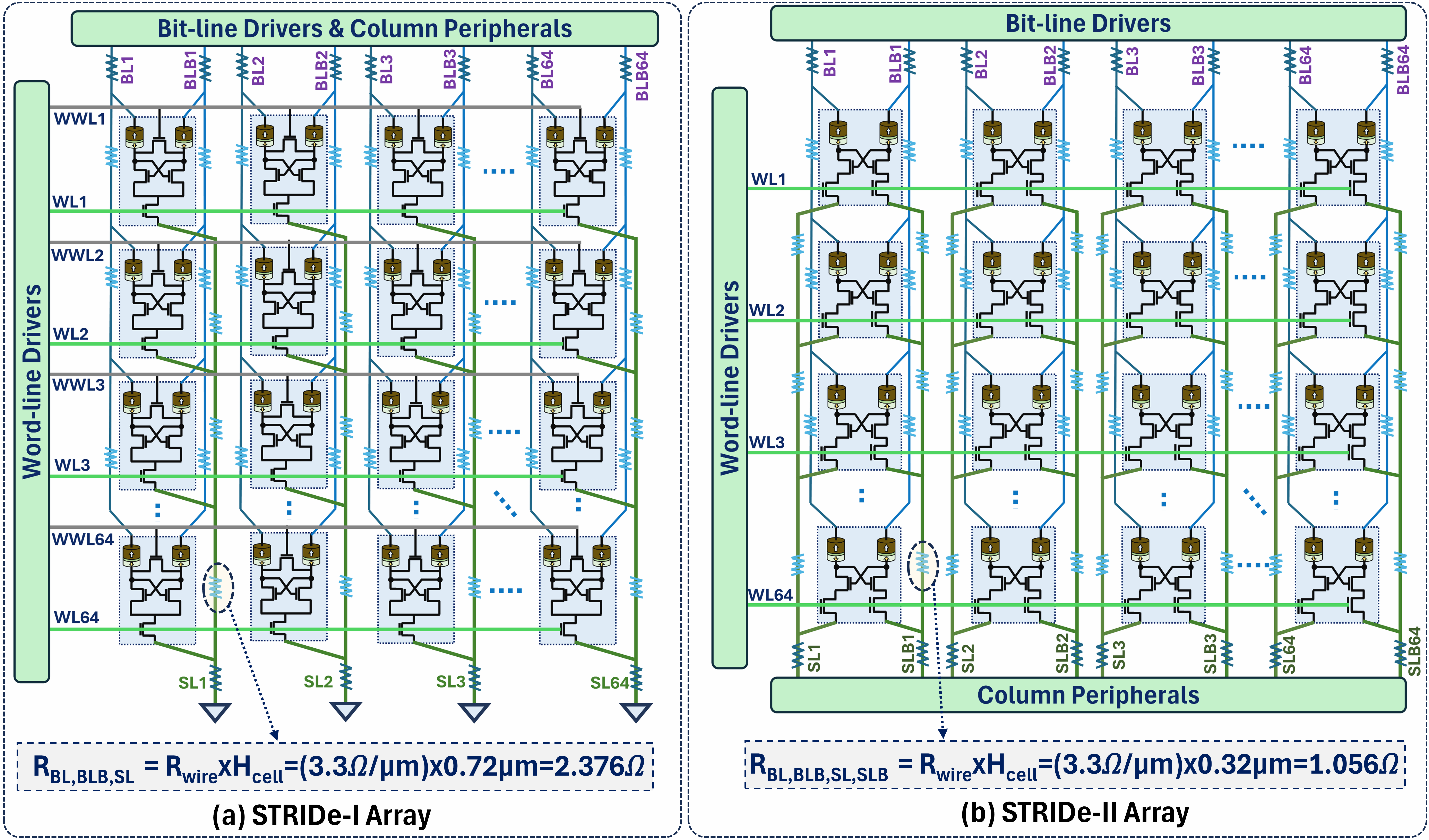}\vspace{-2mm}
\caption{$64\times64$ Crossbar array designs with (a)STRIDe-I and (b)STRIDe-II, including non-ideal resistances. Wire parasitic resistances have been calculated based on resistance per unit-length for 45nm node\cite{moon2008process}. \vspace{-6mm}}
\label{Figure: crossbar}
\end{figure}

There are some interesting points to note here. First, as opposed to standard MRAM designs, high current (I\textsubscript{H}) flows through AP branch and low current (I\textsubscript{L}) flows through P branch in STRIDe bitcells, with significant improvement in the distinguishability between I\textsubscript{H} and I\textsubscript{L}. Second, generally speaking, in standard MRAM bit-cells, read-current-driven STT can potentially and accidentally disturb the MTJ-states. Therefore, operable V\textsubscript{READ} range gets limited in order to prevent read-currents from reaching critical switching-current levels. However, in our design, the direction of read-current is anti-parallelizing, and cross-coupling action makes sure the current in P branch (the anti-parallelizing one) is reduced to a few nA range. Thus, the read-disturb margin of STRIDe is significantly increased, allowing for an increased V\textsubscript{READ} range to operate in. Moreover, for the cross-coupling to be effective, appropriate V\textsubscript{READ} has to be applied to ensure that the cross-coupled transistor on the AP branch is turned ON. This is demostrated in Fig. \ref{Figure: distinguishability_enhancement} as the AP and P branch currents are plotted as a function of V\textsubscript{READ} for STRIDe-I (Fig. \ref{Figure: distinguishability_enhancement}a) and STRIDe-II (Fig. \ref{Figure: distinguishability_enhancement}c) bitcells, with the general trend for both bitcells being the same. Below a certain threshold, both M1 and M2 are OFF with nA range of currents flowing through both branches. However, as V\textsubscript{READ} goes beyond that threshold, the cross-coupling effect starts showing up. With further increase in V\textsubscript{READ}, cross-coupling becomes stronger, opening up a window between I\textsubscript{H} and I\textsubscript{L} with orders of magnitude difference between them. The I\textsubscript{H}/I\textsubscript{L} ratios as a function of V\textsubscript{READ} are shown in Fig. \ref{Figure: distinguishability_enhancement}b,d, showing a maximum I\textsubscript{H}/I\textsubscript{L}=$8156$ for STRIDe-I at V\textsubscript{READ}=$0.66V$ (Fig. \ref{Figure: distinguishability_enhancement}b), and a maximum I\textsubscript{H}/I\textsubscript{L}=$3932$ for STRIDe-II at V\textsubscript{READ}=$0.62V$(Fig. \ref{Figure: distinguishability_enhancement}d). In section \ref{SM}, we will show that even with deviation from the target V\textsubscript{READ} due to IR drops, this distinguishability is still a few thousands, significantly higher than standard MRAMs.

\section{Crossbar Array Design for XNOR-IMC and AND-IMC}\label{crossbar_design}
In this section, we discuss the extraction of parasitic resistances from bitcell layouts, design $64\times64$ crossbar arrays with these non-ideal resistances included, and introduce the encoding schemes for XNOR- and AND-based IMC.\vspace{-2mm}

\subsection{Crossbar Array Design}

Utilizing the bitcell layouts shown in section \ref{bitcells}, we design $64\times64$ STRIDe-I and STRIDe-II crossbar arrays including the driver/wire/sink resistances as shown in Fig. \ref{Figure: crossbar}. The distributed parasitic wire-resistance calculation is according to the technology-specific resistance-per-unit length \cite{moon2008process}. STRIDe-I has larger parasitic wire-resistance per bitcell than STRIDe-II due to its larger bitcell height (Fig. \ref{Figure: bitcells}). During IMC operation, multiple WLs are asserted, BL/BLB are driven to V\textsubscript{READ}, and currents naturally add up on the bit-lines (BL/BLB) and sense-lines (SL/SLB) according to the input-weight combinations of the bitcells, as we will discuss in the next sub-sections. For STRIDe-I, these currents can be sensed from BL and/or BLB, while for STRIDe-II, they can be sensed from SL and/or SLB, as noted earlier.

Note that, the bitcell area of STRIDe-I is $3\times$($1.5\times$) as much as 1T-1MTJ(2T-2MTJ) bitcell, respectively, while the bitcell area of STRIDe-II is $2.67\times$($1.33\times$) that of 1T-1MTJ(2T-2MTJ) bitcell, respectively. However, if we consider the overall IMC-macro area for XNOR-IMC and AND-IMC, the overheads become significantly lower due to the dominance of ADCs and current-subtractors, as we will see in section \ref{hardware}.

\begin{figure}[t!]
\centering
\includegraphics[width=0.45\textwidth]{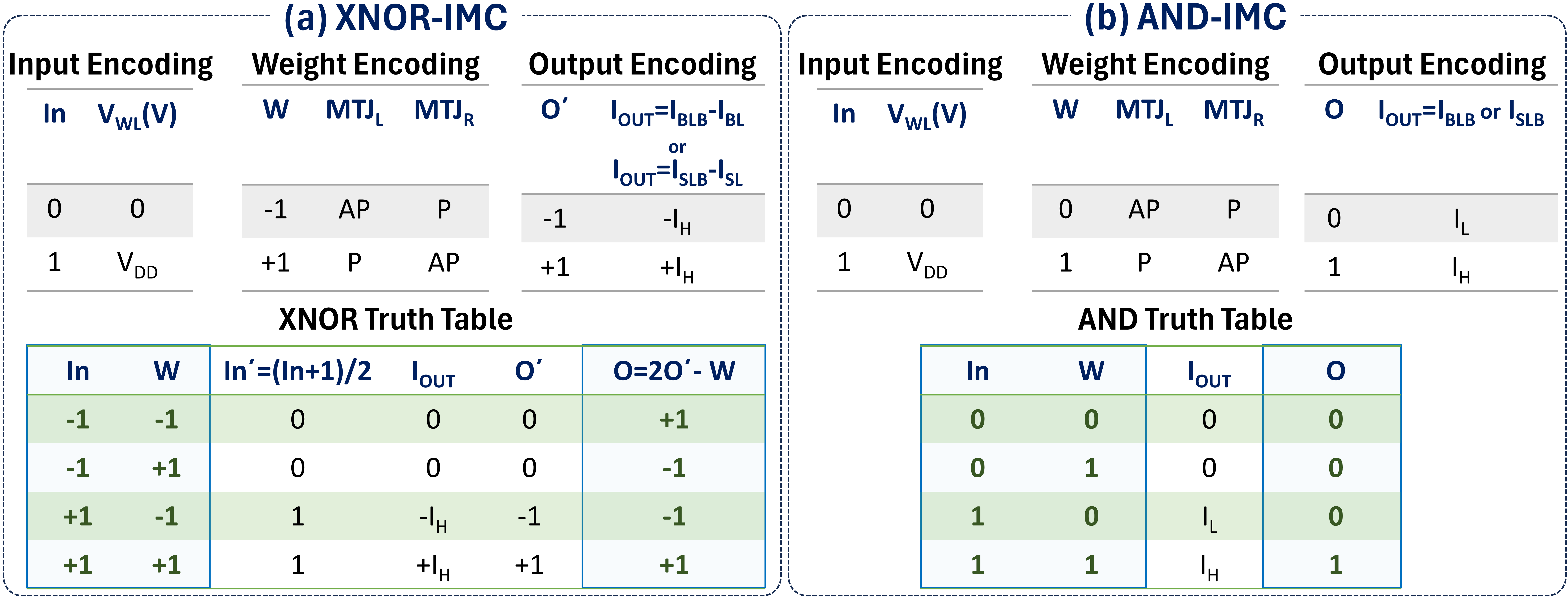}\vspace{-2mm}
\caption{Input, weight and output encoding schemes for (a)XNOR-IMC and (b)AND-IMC with STRIDe.\vspace{-4mm}}
\label{Figure: encoding}
\end{figure}

\subsection{Encoding Scheme for XNOR-IMC}
As we have mentioned before, XNOR-IMC targets MVM for BNNs by performing XNOR operation of \textit{signed} binary inputs and weights, which is equivalent to their scalar multiplication. To implement this, we use the input/weight/output encoding scheme as shown in Fig.\ref{Figure: encoding}a, along with the resulting XNOR truth table. First, the inputs $In$$\in$$\{-1,+1\}$ are transformed into $In'$$\in$$\{0,1\}$ domain using this transformation: \vspace{-1mm}

\begin{equation}
     In'=\frac{1}{2}(In+1)
     \label{eq 2: input conversion}
\end{equation}

This approach is similar to NAND-Net architecture \cite{8675209}. The weights, however, are stored in $W$$\in$$\{-1,+1\}$ domain unlike NAND-Net. Now, the transformed inputs $In'$ are applied to the crossbars as WL voltages, where $In'$=$1$ corresponds to V\textsubscript{WL}=V\textsubscript{DD}=$1.2V$, and $In'$=$0$ corresponds to V\textsubscript{WL}=$0$. As multiple WLs are asserted, bitcell currents naturally accumulate on BL/BLB and SL/SLB. The BL/BLB currents for STRIDe-I and SL/SLB currents for STRIDe-II are passed through analog current-subtractors to get the IMC output, $O'$=$\sum_{k=1}^{n}In'.W$ following the output encoding in Fig. \ref{Figure: encoding}a. This IMC output($O'$) from the crossbar-array column is then digitized using ADC, and the following transformation is applied to extract the XNOR output:\vspace{-1mm}

\begin{equation*}
     \sum_{k=1}^{n}In.W=2\sum_{k=1}^{n}In'.W-\sum_{k=1}^{n}W
     \label{eq 3: XNOR output 1}
\end{equation*}
\vspace{-1mm}
\begin{equation}
     \implies O = 2O'- \sum_{k=1}^{n}W
     \label{eq 3: XNOR output 2}
\end{equation}

Where $O$ is the XNOR output. Multiplying $O'$ by 2 requires a simple left-shift operation. Also, the sum of column-weight ($\sum_{k=1}^{n}W$) can be pre-computed \textit{before} deploying the weights into crossbar arrays, and molded into layer biases, requiring no additional overhead. (Note, in the NAND-Net architecture, both the weights and inputs are transformed to $\{0,1\}$ domain. Thus, for the output post-processing, sum of inputs also needs to be computed, which needs a shared adder tree. The proposed design averts this mild overhead).   

Let us consider a few XNOR examples with an ideal $64\times64$ crossbar array (for now) and PWA of 8, meaning 8 inputs (8 WLs) are asserted in a single cycle. First, let us assume all the 8 inputs are $+1$ and all 8 corresponding bit-cell weights are $+1$ in a column. For STRIDe-I, this yields $I_{BLB}$=$+8I_{H}$, $I_{BL}$$\approx$$0$$\implies$ $I_{OUT}$=$8I_{H}$. For STRIDe-II, this means $I_{SLB}$=$+8I_{H}$, $I_{SL}$$\approx$$0$$\implies$ $I_{OUT}$=$8I_{H}$. 

Now, let us consider all 8 bitcells have $-1$ weights. This means $I_{BLB}$$\approx$$0$, $I_{BL}$=$8I_{H}$$\implies$ $I_{OUT}$=$-8I_{H}$ for STRIDe-I; and $I_{SLB}$$\approx$$0$, $I_{SL}$=$8I_{H}$$\implies$ $I_{OUT}$=$-8I_{H}$ for STRIDe-II. 

For the third example, let us take an arbitrary input pattern, $In$=$[-1,+1,+1,-1,+1,-1,+1,+1]$$\implies$ $In'$= $[0,1,1,0,1,0,1,1]$ and the weight pattern for the 8 bitcells is, $W$=$[-1,+1,-1,-1,+1,-1,+1,+1]$. For STRIDe-I, this leads to $I_{BLB}$=$4I_{H}$ and $I_{BL}$=$I_{H}$ (and for STRIDe-II, $I_{SLB}$=$4I_{H}$, $I_{SL}$=$I_{H}$). The result is $I_{OUT}$=$+3I_{H}$$\implies$$O'$=$ +3$, which is the IMC output in \eqref{eq 3: XNOR output 2}. As the sum of weight is 0 here, from \eqref{eq 3: XNOR output 2} we get, $O$ = $ 2\times(+3)-0$ = $6$, which is the XNOR output for this input-weight combination. To summarize, if m and n bitcells contribute to BLB and BL currents for STRIDe-I, respectively (or SLB and SL currents for STRIDe-II, respectively), $I_{OUT}$=$(m-n)I_{H}$, and $O'$= $m-n$.\vspace{-2mm}

\subsection{Encoding Scheme for AND-IMC}
To implement 4-bit precision DNN inference with STRIDe-I and II, the 4-bit weights are bit-sliced and stored in 4 crossbar-arrays (negative weights are stored in their 2's complement form), while the 4-bit inputs are bit-streamed in 4 cycles as binary WL voltages. Because of the ReLU activation in ResNet18, the inputs are non-negative. Thus, both input and weight bits are in $\{0,1\}$ regime. The MVM of the bit-sliced weights and bit-streamed inputs, therefore, relies on AND-based IMC according to the encoding scheme shown in Fig.\ref{Figure: encoding}b. The MTJs store complementary weights, with MTJ\textsubscript{L} at AP/MTJ\textsubscript{R} at P encoding weight 0 (instead of -1 as for XNOR-IMC), and MTJ\textsubscript{L} at P/MTJ\textsubscript{R} at AP denoting weight $1$ (similar to XNOR-IMC). As multiple WLs are asserted, bitcell currents accumulate on BL and BLB naturally depending on the input-weight pairs. But this time, currents are sensed only at BLB in STRIDe-I, and only at SLB in STRIDe-II. These currents are sent to ADCs to produce the AND-IMC output, $O$. Note, in AND-IMC, no output post-processing is needed.    
For instance, let us consider PWA of 8, assume an input pattern, $In$= $[0,1,1,0,1,0,1,1]$, and a weight pattern, $W$= $[1,1,0,1,0,1,1,0]$. When $In=1$ and $W=0$, the BLB current in STRIDe-I (and SLB current in STRIDe-II) is the low P-branch current, reaching only up to a few nA. Similarly, $In=0$ also results in negligible current irrespective of the weight bit. However, only when both $In=1$ and $W=1$, the BLB current in STRIDe-I (and SLB current in STRIDe-II) is the high AP-branch current in tens of \textmu A range. Hence, for this example, $I_{BLB}$=$2I_{H}$ for STRIDe-I and $I_{SLB}$=$2I_{H}$ for STRIDe-II. This means, $I_{OUT}$=$2I_{H}$$\implies$IMC Output, $O$ = $2$.  

\section{IMC Analysis and Results}\label{imc_results}
In this section, we evaluate the computational robustness of AND-IMC and XNOR-IMC considering $64\times64$ STRIDe crossbar arrays in the presence of circuit non-idealities (driver/wire/sink resistances) and process-variations, and present comparisons against two baseline standard STT-MRAM IMC designs: 1T-1MTJ (with dummy column to improve its robustness by mitigating the impacts of large I\textsubscript{L}) and 2T-2MTJ differential bitcell (with no cross-coupling). Note that, STRIDe can perform both in-memory-AND and in-memory-XNOR operations, according to the encoding schemes shown in Fig. \ref{Figure: encoding}. However, the 1T-1MTJ design inherently can perform only AND-IMC at the crossbar level (the output of which is post-processed in digital domain for conversion into XNOR-output as needed). On the other hand, 2T-2MTJ differential design inherently can perform only XNOR-IMC at the crossbar level (which requires post-processing in digital domain for conversion into AND-output as needed). As we are only focusing on the \textit{crossbar-level} IMC performance evaluation in this section, hence we compare the AND-IMC performance of STRIDe against 1T-1MTJ, and XNOR-IMC performance against 2T-2MTJ.

Before going into the results, let us first clarify the design choices for IMC. The device and circuit parameters used in HSPICE simulations are summarized in Table \ref{table1}. MgO thickness, $t_{MgO}$ has been chosen as $1.3nm$ to reduce MTJ current, and the temperature, T is 25\textdegree C.In the crossbar arrays, SL for STRIDe-I and both SL/SLB for STRIDe-II are biased at virtual ground with the use of op-amps, similar to the design in \cite{8802267}. Moreover, PWA of 8 is used across all the crossbar-designs to- (i) reduce non-ideal effects, and (ii) lower ADC costs (since the maximum absolute IMC output is restricted to 8, we can use 3-bit Flash ADCs). We also investigate their performances under PWA of 16, which has relatively higher IR drops than PWA of 8 and requires 4-bit ADCs, but reduces IMC latency by half (because 4 IMC cycles are required instead of 8). We activate all 64 columns of the asserted rows, maximizing column parallelism. For XNOR-IMC, the sensed currents from BL/BLB in STRIDe-I (and from SL/SLB in STRIDe-II) are passed through a current-subtractor and a comparator as in \cite{thakuria2024sitecimsignedternary} to extract $I_{OUT}$ and the sign of the subtraction result, respectively. However, for AND-IMC, analog current-subtractor is not required, as the currents are sensed only from BLB in STRIDe-I and only from SLB in STRIDe-II.  

Recall that, for AND-IMC, $I_{OUT}$=$I_{BLB}$ in STRIDe-I, and $I_{OUT}$=$I_{SLB}$ in STRIDe-II. For 1T-1MTJ array (performing AND-IMC inherently), $I_{OUT}$=$I_{SL}-I_{SL,dummy}$. On the other hand, for XNOR-IMC, $I_{OUT}$=$I_{BLB}-I_{BL}$ in STRIDe-I, and $I_{OUT}$=$I_{SLB}-I_{SL}$ in STRIDe-II. For 2T-2MTJ differential array (inherently performing XNOR-IMC), $I_{OUT}$=$I_{BL}-I_{BLB}$. (Note, unlike STRIDe, in the standard designs, high current flows in the P branch, low current flows in the AP branch).

One important design aspect we would like to emphasize on is the choice of V\textsubscript{READ}. Instead of choosing the V\textsubscript{READ} where I\textsubscript{H}/I\textsubscript{L} ratio is the maximum, we rather choose a V\textsubscript{READ} slightly greater than the peak-point for both designs. For example, V\textsubscript{READ}=$0.68V$ is chosen for STRIDe-I with I\textsubscript{H}=$22.3\mu A$, I\textsubscript{L}=$2.75nA$, and I\textsubscript{H}/I\textsubscript{L}=$8125$. Similarly, V\textsubscript{READ}=$0.65V$ is chosen for STRIDe-II with I\textsubscript{H}=$21\mu A$, I\textsubscript{L}=$5.52nA$, and I\textsubscript{H}/I\textsubscript{L}=$3800$. Due to non-ideal IR drops, effective bitcell V\textsubscript{READ} is reduced. However, because of such a choice, I\textsubscript{H}/I\textsubscript{L} ratio will increase with the decrease of effective V\textsubscript{READ} at least until the peak-point is reached, unlike the standard MRAM designs. For a fair comparison, we use the same device-parameters across all designs and optimize $V_{READ}$ for the baselines such that their I\textsubscript{H}=$21\mu A$ (similar to I\textsubscript{H} of STRIDe-II).

\begin{figure}[t!]
\centering
\includegraphics[width=0.5\textwidth]{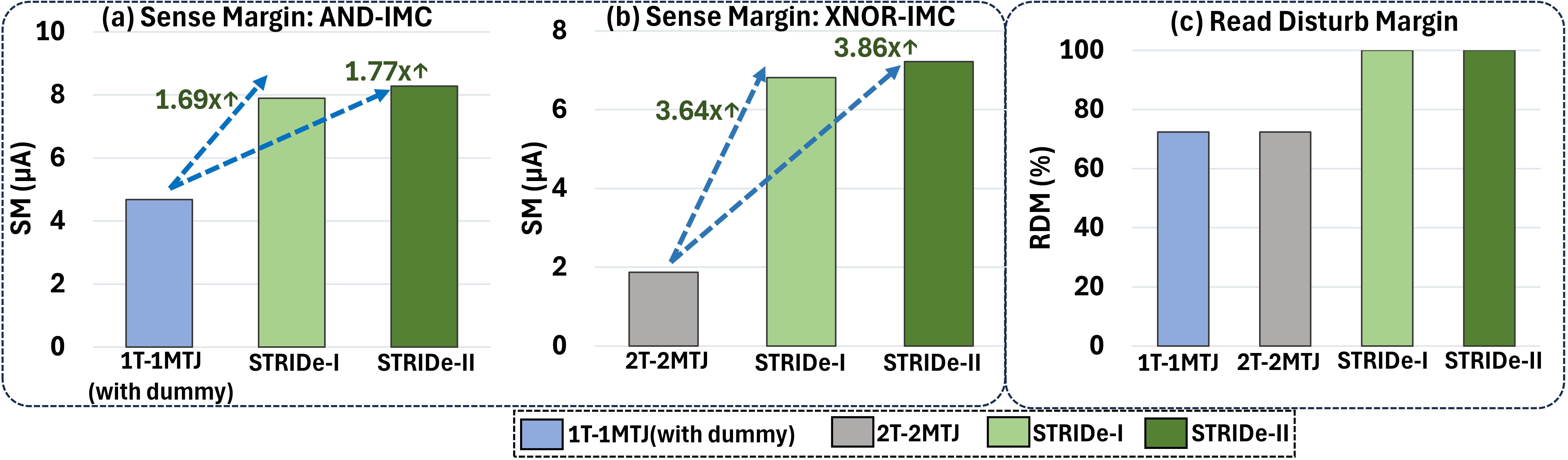}
\caption{IMC robustness comparison of STRIDe designs against standard MRAM designs. (a)Worst-case sense-margin (SM) comparison against 1T-1MTJ (with dummy) for AND-IMC, (b)Worst-case SM comparison against 2T-2MTJ for XNOR-IMC, (c) Read disturb margin comparison.\vspace{-4mm}}
\label{Figure: IMC_results}
\end{figure}

\subsection{Sense Margin (SM) Analysis} \label{SM}
Sense margin (SM) is an important IMC-robustness metric, which quantifies the distinguishability between neighboring output states and is defined as:

\begin{equation}
    \textup{SM}= \frac{I_{OUT,a|min}-I_{OUT,a-1|max}}{2}
    \label{eq 4: SM}
\end{equation}

Here, $I_{OUT,a|min}$ is the minimum output-current corresponding to an output state $a$, and $I_{OUT,a-1|max}$ is the maximum output-current corresponding to the preceding output state $a-1$. Different input-weight combinations in a crossbar array may correspond to the same column IMC-output. But due to their relative positions in the array, these combinations face different non-ideal IR drops, resulting in different output currents for the same IMC-output state. As a result, one specific output state is mapped to a range of non-ideal output currents instead of getting mapped to a single ideal-current. Under this scenario, the chances of overlaps between neighboring output states increases, thereby decreasing SM and increasing compute error probability.

To examine this, we apply 8000 different input-weight combinations to simulate the crossbar arrays (with PWA = 8), perform both AND- and XNOR-based IMC, extract I\textsubscript{OUT} from each column to calculate SM, and then obtain the minimum SM value. Our results show that the worst-case SM for AND-IMC with STRIDe-I is $7.89\mu A$ and with STRIDe-II is $8.28\mu A$. These are $1.69\times$ and $1.77\times$ higher, respectively, compared to 1T-1MTJ with a dummy column ($4.68\mu A$, Fig. \ref{Figure: IMC_results}a). We also see significant SM improvement in XNOR-IMC, as the worst-case SM for STRIDe-I is $6.81\mu A$ and for STRIDe-II is $7.22\mu A$- a $3.64\times$ and $3.86\times$ improvement over 2T-2MTJ differential design (with a worst case SM=$1.87\mu A$), respectively (Fig. \ref{Figure: IMC_results}b). Note that, these simulation results are under nominal conditions, i.e., without process-variations. However, this SM enhancement of STRIDe eventually helps achieve process-variation tolerance as we will discuss shortly.  

The SM improvements of STRIDe designs result from the significant bitcell-level distinguishability enhancement. To understand this, let us look at the impacts of IR drops in our design. At our chosen V\textsubscript{READ}, I\textsubscript{H}/I\textsubscript{L} ratio is $8125$ for STRIDe-I and $3800$ for STRIDe-II. Now, due to the non-ideal resistances, BL and BLB terminals of the STRIDe crossbar arrays face unwanted drops in V\textsubscript{READ} based on input-weight-dependence. The \textit{worst} effective V\textsubscript{READ} (or the lower bound of V\textsubscript{READ}) that can appear at the BL/BLB of a bitcell can be estimated as:\vspace{-3mm}

\begin{equation*}
V_{min,eff}=V_{READ}-I_{H}.pwa.\{R_{D}+(N-pwa-1)R_{w}\} 
\end{equation*}\vspace{-3mm}
\begin{equation}
-I_{H}R_{w}\sum_{k=1}^{pwa}k
\label{eq 5: Vmineff}
\end{equation}

Where, $pwa$=number of wordlines asserted,$R_{D}$=driver resistance,$R_{w}$=distributed wire resistance. In this worst-case estimation, we consider that: (i) all the asserted wordlines have V\textsubscript{WL}=$1.2V$, (ii) all the corresponding bitcells draw I\textsubscript{H} each through either BL or BLB (worst case), and (iii) these bitcells are at the bottom of the crossbar array for the IR drops to be the most severe. Under these assumptions, the worst possible effective V\textsubscript{READ} for STRIDe-I is $0.61V$, with I\textsubscript{H}/I\textsubscript{L}=$7800$  and for STRIDe-II it is $0.59V$ with I\textsubscript{H}/I\textsubscript{L}=$3876$. Therefore, even under IR drops, the distinguishability is still large, and the designs operate in the safe (high $I_H/I_L$) region.

Another benefit of STRIDe is that, the low current is dragged down to a few nA, which significantly reduces the IR drops in the associated branches. In contrast, I\textsubscript{LOW} for the baseline designs is $\approx 3.87\mu A$, which is orders of magnitude larger compared to STRIDe-bitcells, severely degrading IR drops and causing more deviation from ideal-currents. 

The enhanced distinguishability of STRIDe designs coupled with reduced impact of IR drops allows for turning on more than 8 wordlines per IMC-cycle while maintaining high SM. Now, applying PWA of 16 (asserting 16 wordlines in one IMC-cycle) helps reduce overall IMC latency, but it causes larger currents to accumulate on the bitlines and sense-lines, increasing IR drops and deteriorating SM in general. To verify how this impacts the SM of the STRIDe designs and the baselines, we apply PWA = 16 for the same 8000 input-weight combinations. Our results show that, 1T-1MTJ and 2T-2MTJ crossbars suffer significantly due to the increased IR drops, as the worst-case SM for both of them becomes negative (due to output current overlaps between neighboring states). However, STRIDe-I maintains a worst-case SM of $0.74\mu A$ for XNOR-IMC and $3.8\mu A$ for AND-IMC, whereas for STRIDe-II, these values are $2.75\mu A$ for XNOR-IMC and $6.9\mu A$ for AND-IMC. Thus, STRIDe designs maintain notable IMC-robustness even with PWA of 16, while the baseline designs suffer. The implications of this will be discussed in section \ref{inference_accuracy} where we present the inference accuracies with these designs.

\begin{figure}[t!]
\centering
\includegraphics[width=0.5\textwidth]{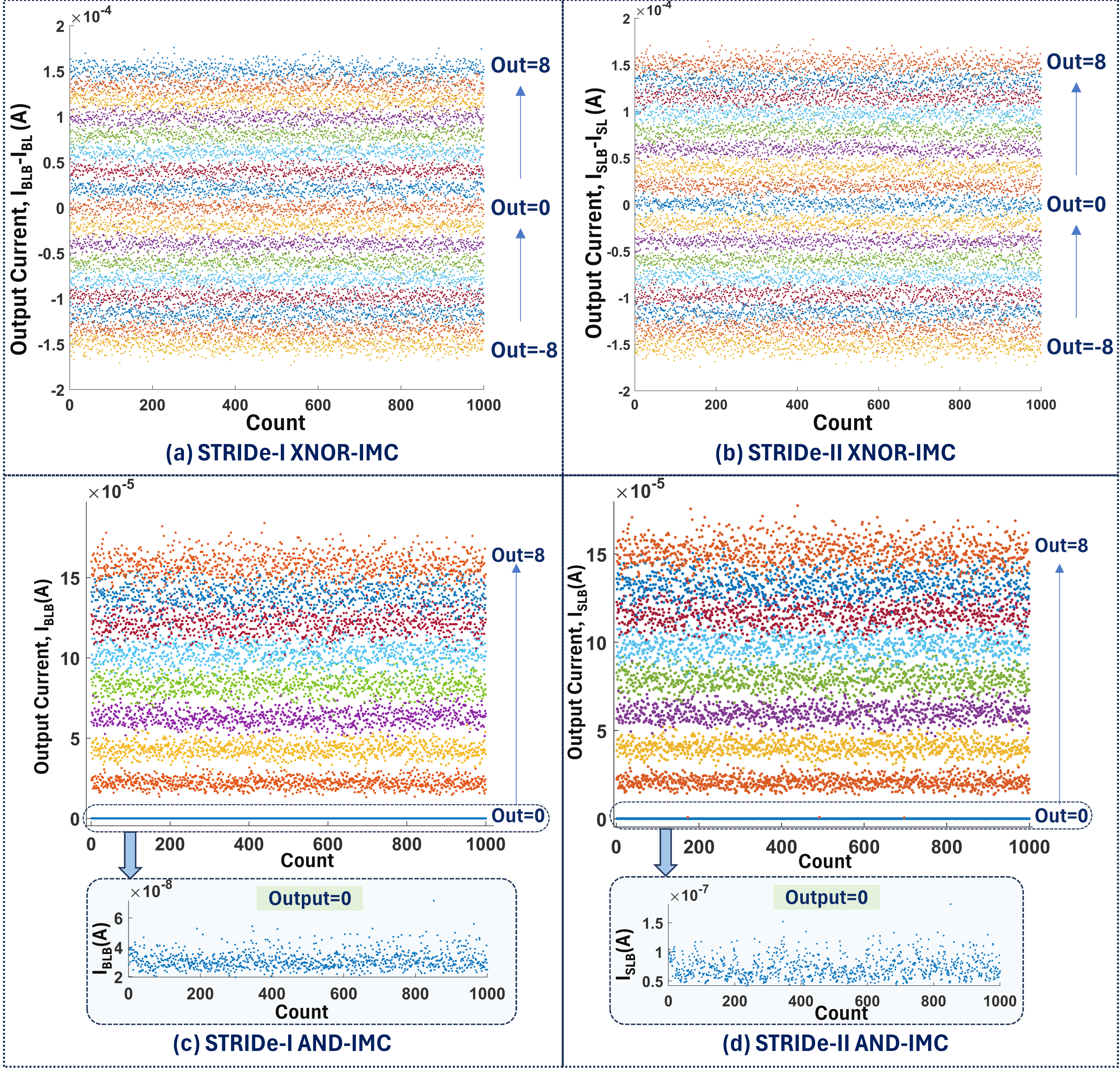}\vspace{-2mm}
\caption{Results for 1000 Monte Carlo simulations per output state with STRIDe-I and II under PWA of 8. (a)XNOR-IMC with STRIDe-I, (b)XNOR-IMC with STRIDe-II, (c)AND-IMC with STRIDe-I (inset: output currents for output=0), (d)AND-IMC with STRIDe-II (inset: output currents for output=0).\vspace{-4mm}}
\label{Figure: processvariation}
\end{figure}

\subsection{Read Disturb Margin (RDM) Analysis}
It is important to ensure that during the read/IMC operation, the MTJ weights are retained and do not get accidentally switched. Read disturb margin (RDM) is a metric that quantifies the robustness to this possibility of accidental read-disturb, which is given by:

\begin{equation}
\textup{RDM}=\frac{I_{CR}-I_{MTJ}}{I_{CR}}\times 100\%
\label{eq 6: RDM}
\end{equation}

Here, $I_{CR}$=critical switching current, and $I_{MTJ}$= actual MTJ current. Operating closer to $I_{CR}$ increases the probability of read disturb. As the read current direction in our designs is anti-parallelizing, we only consider the critical current for P\textrightarrow AP switching, which is $I_{CR}$=$75.96\mu A$. As I\textsubscript{P}=$21\mu A$ for the two baselines, RDM for both of them is $72.35\%$, whereas for STRIDE-I and II bitcells, RDM values are $99.996\%$ and $99.992\%$, respectively (Fig. \ref{Figure: IMC_results}c). This $27.6\%$ RDM boost is the result of cross-coupling action drastically reducing I\textsubscript{P} down to just $2.75nA$ and $5.52nA$ for STRIDe-I and II bitcells, respectively.

\subsection{Process-Induced Variations}\label{montecarlo}
To demonstrate how STRIDe crossbar arrays perform under process-variations, we carry out 1000 Monte Carlo (MC) simulations per output state for STRIDe-I and STRIDe-II designs under PWA of 8 considering the following variations: (i) Standard deviation ($\sigma$) of transistor threshold voltage ($V_{th}$) = $25mV$, (ii) $\sigma$ of MTJ oxide thickness, $t_{MgO}$ = $1.5\%$, and (iii) $\sigma$ of MTJ diameter = $5\%$ of minimum metal width (which is 65nm for 45nm technology node)\cite{10268108}. The simulation results are shown in Fig. \ref{Figure: processvariation}. In general, owing to the enhanced SM of the STRIDe designs, the output currents have more room to spread out and deviate from ideal values before overlapping with the currents corresponding to neighboring output states. This is true especially for the lower IMC outputs (the most frequent ones) for ResNet18 BNN and 4-bit DNN inference on CIFAR10 dataset. Note that, the spread of output currents for XNOR-IMC near lower \textit{unsigned} outputs (Fig. \ref{Figure: processvariation}a,b) are wider than for AND-IMC(Fig. \ref{Figure: processvariation}c,d). This is because, for XNOR-IMC, the number of input-weight combinations which may result in a specific output is much larger compared to the number of combinations resulting in the same output for AND-IMC, especially near lower outputs. For example, an output of 0 for XNOR-IMC may occur whenever BL and BLB/SL and SLB carry the same output currents, which is possible across multiple different input-weight combinations. Based on the output current values, the IR drops vary significantly across these combinations. This results in a wider spread in the subtracted currents (e.g., output currents). In contrast, for AND-IMC, just one of the input/weight bits being 0 is enough for an output to be 0, the resulting output current in all these cases stays in nA range even with variations. This results in a reduced spread for lower outputs. Nevertheless, the improved SM of STRIDe increases room for these spreads, translating to higher inference accuracies (discussed later).

\begin{figure}[t!]
\centering
\includegraphics[width=0.5\textwidth]{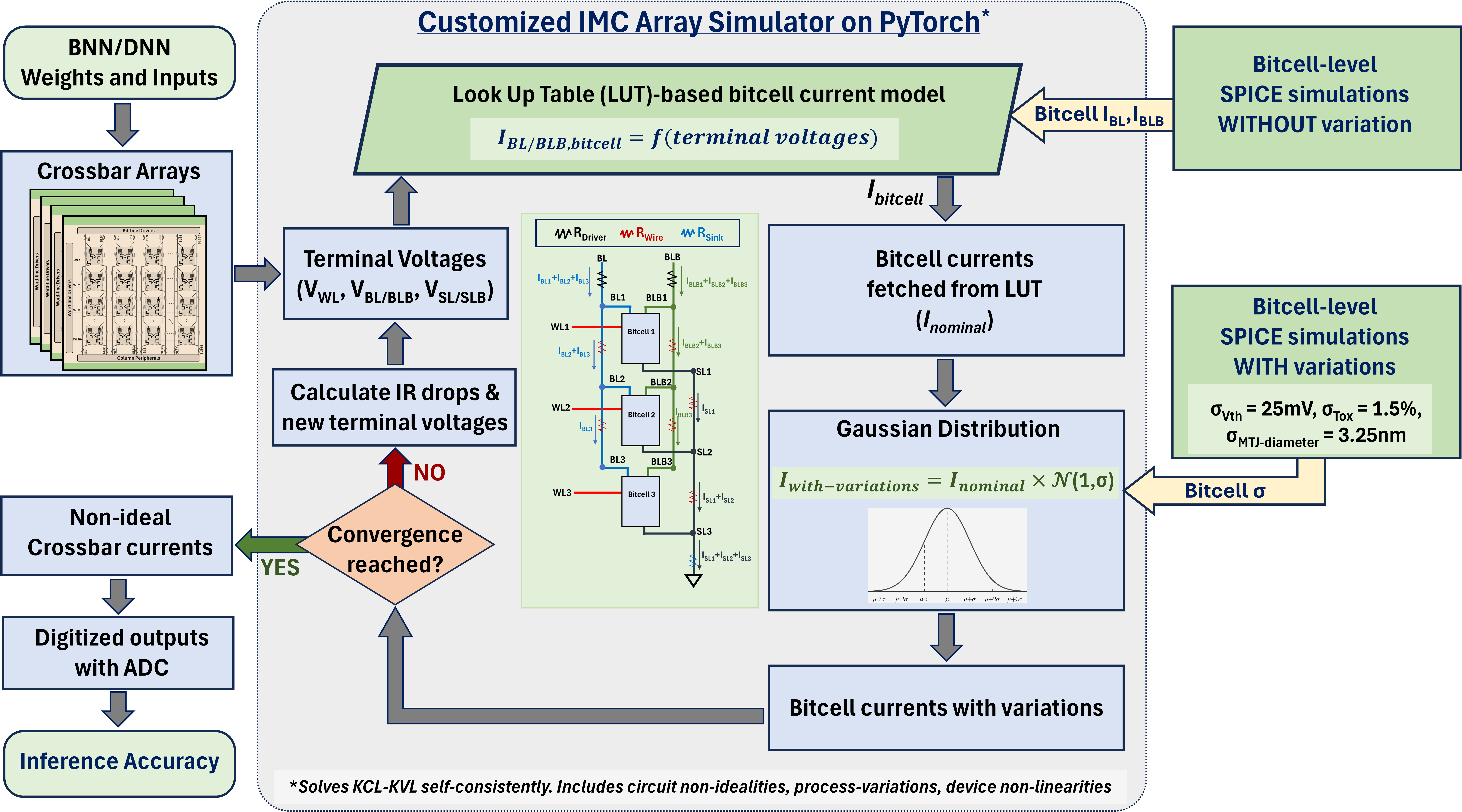}\vspace{-2mm}
\caption{Simulation framework for BNN and 4-bit DNN inference with PyTorch-based customized IMC-array solver.\vspace{-4mm}}
\label{Figure: framework}
\end{figure}

\section{System Level Performance Evaluation}\label{system_level_results}
In this section, we deploy ResNet18 BNN and 4-bit (weight and input precision) DNN inference workloads trained on CIFAR10 dataset on the four crossbar array designs and present inference accuracy comparisons among them under both PWA of 8 and 16. We also discuss the overall macro-level IMC energy-latency-area overheads incurred by each of these designs.

\subsection{Evaluation Framework}
For rigorously evaluating the inference accuracy with the two STRIDe designs and the two baseline MRAM designs, we develop a customized PyTorch-based crossbar array solver\textemdash a simulation platform that allows for seamless incorporation of hardware non-idealities into DNN workflow (Fig. \ref{Figure: framework}). The simulator is similar to the one in \cite{11008637}, which self-consistently solves Kirchoff's voltage/current law (KVL/KCL), taking into account hardware non-idealities (driver/source/sink resistances) and device non-linearities (by forming bit-cell current look-up tables or LUTs as a function of the bitcell terminal voltages). The LUTs for STRIDe bitcells and the baseline bitcells are obtained from HSPICE simulations. During DNN inference, the weights are mapped onto multiple crossbar arrays and input bits are applied as binary WL voltages for BNN, or streamed in 4-cycles for 4-bit DNN. These arrays are then solved using the simulator using an iterative approach. At each iteration, the simulator calculates the terminal voltages for each bitcell in the array, fetches the corresponding bitcell currents from LUTs as a function of these terminal voltages, and, at the next iteration, recalculates the terminal voltages using the fetched currents accounting for the IR drops. This cycle continues until convergence is reached. Our simulator has been validated against HSPICE simulations, with a considerable tool-accuracy showing a maximum error of $<0.3\%$.   

Additionally, we employ a Gaussian distribution on the bitcell currents to model variations in our framework. For this, we use the same variations as described in section \ref{montecarlo} and perform 1000 Monte Carlo simulations on each bitcell. Then we extract the standard deviation ($\sigma$) values from the resulting output current distributions and incorporate them into the bitcell currents using the following to account for variations:\vspace{-3mm}

\begin{equation}
I_{with-variations} = I_{nominal}*\mathcal{N}(1,\sigma)
\label{eq 7: variation}
\end{equation}
\vspace{-3mm}

In case of the baseline bitcells, the extracted $\sigma$ values (at the \textit{bitcell} level) are approximately $16\%$ for P branch and $17.4\%$ for AP branch. In contrast, the $\sigma$ values for the AP branch (I\textsubscript{H}) of STRIDe-I and II are approximately $12\%$ and $15.5\%$. Interestingly for P branch (I\textsubscript{L} driven by OFF transistors), these values are approximately $76.37\%$ and $77.83\%$ for STRIDe-I and II, respectively. Although it seems like a large variation, it is important to note that the P branches of STRIDe are operating in the OFF state. Hence, even with this much variation, I\textsubscript{L} is still in nA range and has minimal deteriorating impact on IMC robustness, helping the STRIDe designs maintain high inference accuracies under variation.  

Thus, the simulator accurately calculates the IMC-array currents, accounting for crossbar parasitic resistances, device non-linearities and process-variations. These currents are then digitized with linear ADC reference-levels to extract the non-ideal IMC outputs. Our PyTorch-based crossbar array solver is directly integrated within the inference accuracy simulations so that the inference accuracy is obtained using the non-ideal IMC outputs. Due to such a direct integration of the non-ideal crossbar model, the non-ideal IMC outputs correspond to the actual input/weight bits encountered during the inference flow.

\begin{figure}[t!]
\centering
\includegraphics[width=0.5\textwidth]{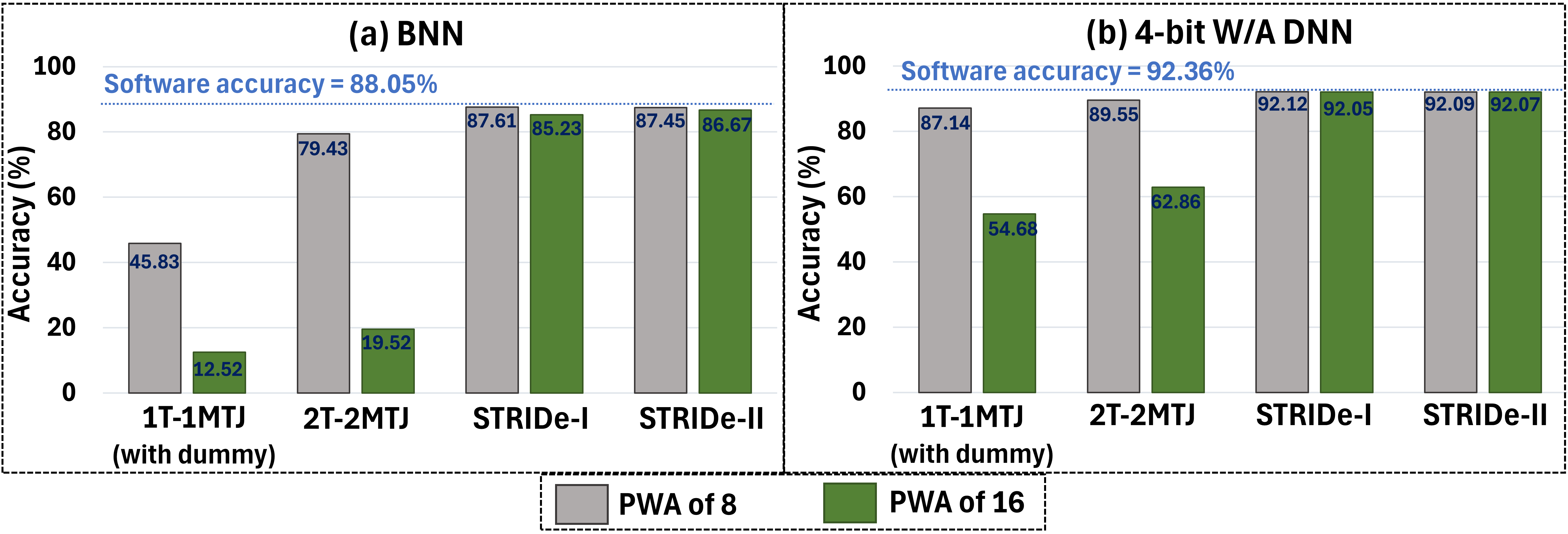}\vspace{-2mm}
\caption{Inference accuracies for (a)ResNet18 BNN, and (b)4bit weight/input ResNet18 DNN on CIFAR10 for the baseline IMC designs and the STRIDe designs. STRIDe-I and II achieve near software accuracies with both PWA of 8 and 16.\vspace{-4mm}}
\label{Figure: accuracy}
\end{figure}

\subsection{Inference Accuracy Evaluation}\label{inference_accuracy}
Recall that, the actual output currents of the crossbar arrays deviate from ideal current values due to the non-ideal IR drops. This increases inaccuracies in IMC outputs if we digitize using the ideal ADC reference levels (with reference quantization current, I\textsubscript{quant} = I\textsubscript{OUT,bitcell}), as it does not account for the deviation. To minimize this effect and get the best inference accuracies for all the designs, we employ ADC reference current optimization, where we use ADC levels with \textit{lower} reference current-level (I\textsubscript{quant}) which minimizes the errors introduced by the non-ideal deviation of output currents on average. This approach is similar to the one in \cite{11008637}. Additionally, PWA of 8 or 16 is applied to further mitigate the impacts of non-idealities. 

Let us first analyze the ResNet18 BNN inference accuracies on CIFAR10 dataset, summarized in Fig. \ref{Figure: accuracy}a. The software accuracy stands at $88.05\%$. Now, 1T-1MTJ with no dummy column yields a poor accuracy of only $9.98\%$ for PWA of 8. Hence, as we mentioned earlier, we use dummy column as a mitigation technique, and also apply ADC level optimization and PWA of 8, resulting in an accuracy of $45.83\%$, which is still not satisfactory. The 2T-2MTJ structure shows improved robustness due to its differential nature and recovers accuracy up to $79.43\%$. In contrast, the enhanced SM (along with process-variation tolerance) of STRIDe-I and II translates to near-software-accuracies of $87.61\%$ and $87.45\%$ respectively. These are just $0.44\%$ and $0.60\%$ degradations from software accuracy. The slightly higher accuracy of STRIDe-I than STRIDe-II comes from its higher distinguishability (higher I\textsubscript{H}/I\textsubscript{L}) and lower I\textsubscript{L}. As we apply PWA of 16, the IR drops increase, and the accuracies of both 1T-1MTJ and 2T-2MTJ degrade significantly to $12.52\%$ and $19.52\%$, respectively. However, STRIDe-I and STRIDe-II maintain $85.23\%$ and $86.67\%$ accuracies respectively, even under PWA of 16. The slightly larger accuracy of STRIDe-II compared to STRIDe-I is due to its smaller wire resistances in the vertical direction compared to STRIDe-I (due to lower layout height-see Fig. \ref{Figure: crossbar}), which helps reduce the impact of IR drops. \vspace{-0.5mm}

\begin{figure*}[t!]
\centering
\includegraphics[width=0.9\textwidth]{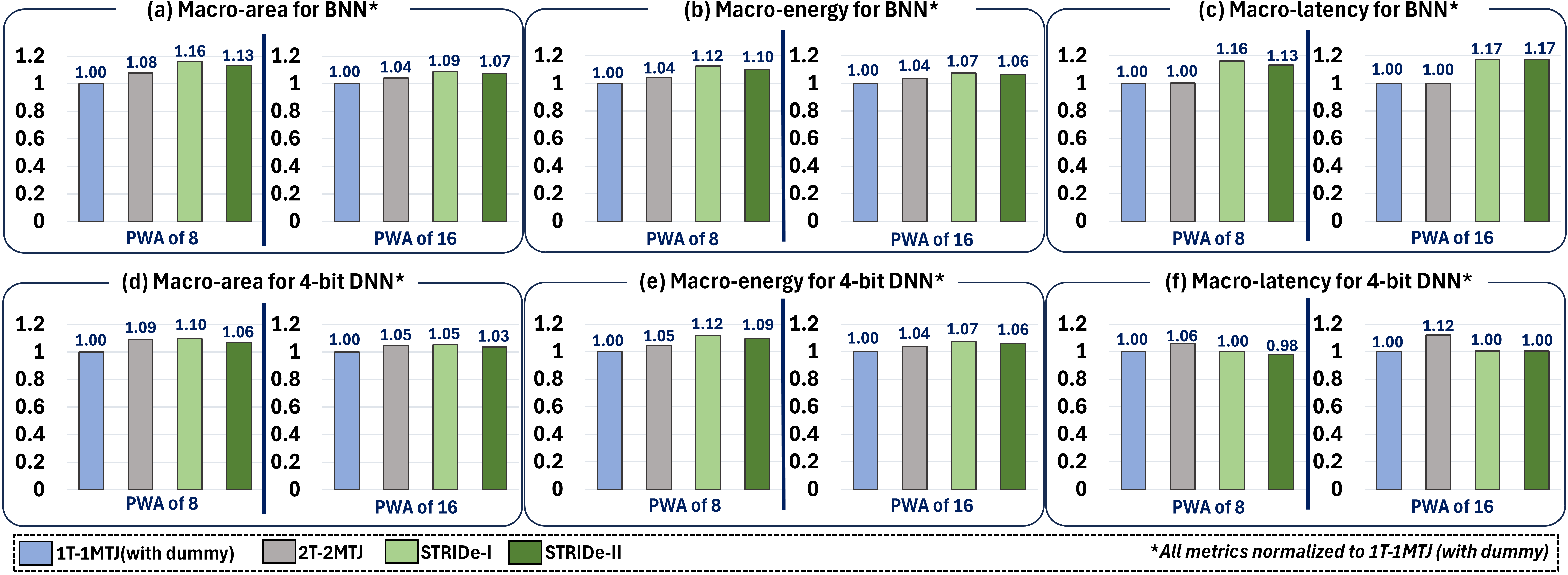}\vspace{-2mm}
\caption{IMC macro-level energy-latency-area comparisons for BNN and 4-bit DNN inference. All metrics are normalized to 1T-1MTJ design with dummy column.\vspace{-5mm}}
\label{Figure: overheads}
\end{figure*}

Now, let us analyze their inference accuracies for DNNs with 4-bit weights and inputs, as shown in Fig. \ref{Figure: accuracy}b. As discussed before, the weights are bit-sliced and stored on 4 crossbars (negative weights stored in 2's complement form), while the 4 input bits are streamed in 4 cycles. The software inference accuracy for our 4-bit ResNet18 DNN trained on CIFAR10 is $92.36\%$. Under PWA of 8, 1T-1MTJ without dummy column yields only $18.75\%$ of accuracy, while 1T-1MTJ with dummy recovers it to $87.14\%$. 2T-2MTJ results in inference accuracy of $89.55\%$. In contrast, STRIDe-I and II achieve $92.12\%$ and $92.09\%$ accuracies, respectively. Note that, 1T-1MTJ (with dummy) and 2T-2MTJ designs perform reasonably well for PWA of 8 in this case, which results mainly from the higher sparsity of 4-bit DNNs compared to BNNs. Lower sparsity of BNN weight and input profiles tend to generate larger MVM outputs on average. As non-ideality induced current deviation depends superlinearly as a function of MVM output (as shown in \cite{11008637}), hardware implementation of BNN inference, in general, is significantly susceptible to non-idealities. In contrast, more sparse weight and input profiles of higher precision networks like 4-bit DNNs tend to generate lower MVM outputs on average, thereby reducing non-ideal impacts. This, coupled with PWA of 8, benefits all four designs. As we turn on more wordlines with PWA of 16, the inference accuracies for 1T-1MTJ and 2T-2MTJ drop down to $54.68\%$ and $62.86\%$, whereas STRIDe-I and II maintain $92.05\%$ and $92.07\%$, respectively. As the enhanced distinguishability of STRIDe designs allows for turning on larger number of wordlines simultaneously while maintaining near-ideal inference accuracies, it offers a higher design flexibility. In particular, one has the option to reduce the overall system-level latency by turning on more WLs, albeit with the requirements of higher ADC bit-precision.

\subsection{Hardware Analysis: Energy-Latency-Area}\label{hardware}
Fig. \ref{Figure: overheads} summarizes the macro-level energy-latency-area overheads incurred by the two STRIDe designs compared to the baselines for both BNN and 4-bit DNN inference workloads. For BNN accelerators, all the designs require wordline decoders/drivers (for PWA of 8 and 16), current subtractors (1T-1MTJ for dummy current subtraction, others for getting XNOR output from IMC array), and ADCs for converting analog output currents to digital MVM outputs (3-bit and 4-bit flash ADCs for PWA of 8 and 16, respectively). In addition, 1T-1MTJ requires adder trees to post-process and convert the IMC-output back to XNOR output. As mentioned before, the cost of these adder trees can be amortized by sharing them across multiple crossbars. 

With these peripherals considered for PWA of 8, STRIDe-I comes with a $16.3\%$($7.7\%$) macro area overhead over 1T-1MTJ (2T-2MTJ) baseline (Fig. \ref{Figure: overheads}a). This overhead is much less (compared to the bit-cell area) because IMC array constitutes a small fraction of the overall macro, with the subtractors and ADCs being the dominant components. As for IMC energy, STRIDe-I incurs $12.6\%$($7.85\%$) overhead over 1T-1MTJ (2T-2MTJ) designs (over 8 cycles for PWA of 8), with ADCs and subtractors dominating the energy consumption (Fig. \ref{Figure: overheads}b). Finally, there is a $16.3\%$ ($16\%$) larger IMC-macro latency introduced by STRIDe-I compared to 1T-1MTJ (2T-2MTJ) (Fig. \ref{Figure: overheads}c). It takes slightly longer for the STRIDe-I output currents to reach steady-state compared to the baselines due to the cross-coupling action, hence this increase.

Now, STRIDe-II IMC-macro takes up $13.4\%$($5.27\%$) larger area than 1T-1MTJ (2T-2MTJ) IMC macro (Fig. \ref{Figure: overheads}a). The smaller area overhead compared to STRIDe-I design results from the lower bitcell/crossbar-array area of STRIDe-II than that of STRIDe-I. The macro energy overhead incurred by STRIDe-II is $10.22\%$ ($5.5\%$) over 1T-1MTJ (2T-2MTJ), with energy consumption by the ADCs and the subtractors being the dominant components (Fig. \ref{Figure: overheads}b). As for latency, STRIDe-II macro has a $13.3\%$ ($13.07\%$) overhead compared to 1T-1MTJ (2T-2MTJ) design (Fig. \ref{Figure: overheads}c).

For 4-bit DNN, wordline decoders are required across all designs. Area-heavy current subtractors are needed for both 1T-1MTJ (for dummy current subtraction) and 2T-2MTJ (for getting XNOR-IMC output). Further, 2T-2MTJ needs to convert the XNOR-outputs to AND-output, which requires sum of input ($\sum In$) computation using adder trees. This adder tree requirement increases the overheads of 2T-2MTJ. In contrast, for STRIDe-I(STRIDe-II), we read out $I_{BLB}$($I_{SLB}$)  which correspond to the AND-output. Hence, they \textit{do not} require the subtractors and other post-processing circuitry. 3-bit (or 4-bit) flash ADCs are required for all designs under PWA of 8 (or 16). The results for PWA of 8 show that, STRIDe-I incurs $9.58\%$ ($0.51\%$) macro-area overhead over 1T-1MTJ (2T-2MTJ) design (Fig. \ref{Figure: overheads}d). The adder tree requirement of 2T-2MTJ increases its macro area, making it comparable to the area of STRIDe-I macro. Next, the energy overhead of STRIDe-I is $11.9\%$ ($7.19\%$) over 1T-1MTJ (2T-2MTJ) design(Fig. \ref{Figure: overheads}e). 

Although STRIDe-II takes up $6.73\%$ larger area compared to 1T-1MTJ, it actually requires $\sim$$2\%$ \textit{lower} area compared to 2T-2MTJ(Fig. \ref{Figure: overheads}d). This reduction in overall area is a combined effect of the adder tree requirement of 2T-2MTJ and the lower bitcell/IMC-array area of STRIDe-II (compared to STRIDe-I). The macro energy overhead of STRIDe-II is $9.57\%$ ($4.85\%$) over 1T-1MTJ (2T-2MTJ) designs (Fig. \ref{Figure: overheads}e). Interestingly, both the STRIDe designs incur similar IMC latency compared to 1T-1MTJ, and $\sim$$5\%$ \textit{lower} latency compared to 2T-2MTJ (Fig. \ref{Figure: overheads}f). As the 2T-2MTJ macro requires subtractors, ADCs, and adder trees\textemdash all being in the critical path\textemdash it has the highest latency of the four designs.

Besides PWA of 8, we also present the energy-latency-area comparisons for PWA of 16 in Fig. \ref{Figure: overheads}. These results follow similar trends as for PWA of 8, with PWA of 16 reducing overall latencies for all designs as discussed before. However, it should be noted that both baselines yield poor inference accuracies under PWA of 16 for ResNet18 BNN and 4-bit DNN, while the STRIDe designs maintain significantly higher inference accuracies.

\section{Conclusion}
To summarize, we attack the fundamental problem of low distinguishability in STT-MRAM at the very bitcell level and propose STRIDe-I and II for robust XNOR- and AND-based MRAM-IMC for deep neural networks. The two proposed MRAM designs utilize cross-coupling action of two MTJ branches per bitcell to significantly reduce I\textsubscript{L} and enhance I\textsubscript{H}/I\textsubscript{L} ratio \textit{in the bitcell}. As a result, the impact of hardware non-idealities during IMC is significantly reduced, leading to enhanced sense margin and paving way for robust IMC. This robustness improvement translates to the achievement of close-to-software inference accuracies with STRIDe designs for both ResNet18 BNN and 4-bit DNN (trained on CIFAR10) compared to the baseline STT-MRAM IMC designs. Moreover, the distinguishability enhancement of STRIDe at the bitcell level allows for turning on higher number of wordlines (compared to the baselines) while maintaining acceptable inference accuracies, thereby enabling higher row parallelism and reduction in computation latency. These benefits come with some energy-latency-area costs.

\bibliographystyle{IEEEtran}
\bibliography{reference.bib}

\vfill

\end{document}